\documentclass[a4paper,11pt]{article}

\usepackage{a4wide}
\usepackage[english]{babel}

\usepackage{amsfonts}
\usepackage{amsmath}
\usepackage{graphicx}
\usepackage{caption}
\usepackage[colorlinks,bookmarks=true]{hyperref}
\usepackage{amsthm}
\usepackage{enumitem}

\newtheorem{lemma}{Lemma}[section]

\newtheorem{theorem}{Theorem}[section]
\newtheorem{prop}{Proposition}[section]
\newtheorem{remark}{Remark}[section]

\newtheorem{assumption}{Assumption}[section]

\newtheorem{corollary}{Corollary}[section]

\newcommand{\E}{\mathbb{E}}

\title{The microstructural foundations of leverage effect and\\ rough volatility}

\author{Omar El Euch\\ CMAP, \'Ecole Polytechnique Paris \\ omar.el-euch@polytechnique.edu\\$~~$\\
Masaaki Fukasawa\\ 
Graduate School of Engineering Science, Osaka University\\ fukasawa@math.sci.osaka-u.ac.jp\\$~~$\\
Mathieu Rosenbaum\\ CMAP, \'Ecole Polytechnique Paris\\
  mathieu.rosenbaum@polytechnique.edu}

\begin{document}

\maketitle

\begin{abstract}
\noindent We show that typical behaviors of market participants at the high frequency scale generate leverage effect and rough volatility. To do so, we build a simple microscopic model for the price of an asset based on Hawkes processes. We encode in this model some of the main features of market microstructure in the context of high frequency trading: high degree of endogeneity of market, no-arbitrage property, buying/selling asymmetry and presence of metaorders. We prove that when the first three of these stylized facts are considered within the framework of our microscopic model, it behaves in the long run as a Heston stochastic volatility model, where leverage effect is generated. Adding the last property enables us to obtain a rough Heston model in the limit, exhibiting both leverage effect and rough volatility. Hence we show that at least part of the foundations of leverage effect and rough volatility can be found in the microstructure of the asset.
\end{abstract}

\noindent \textbf{Keywords:} Market microstructure, high frequency trading, leverage effect, rough volatility, Hawkes processes, limit theorems, Heston model, rough Heston model.

\section{Introduction}\label{intro}

Leverage effect is a well-known stylized fact of financial data. It refers to the negative correlation between price returns and volatility increments: when the price of an asset is increasing,
its volatility drops, while when it decreases, the volatility tends to become larger.
The name ``leverage" comes from the following interpretation of this phenomenon due to Black \cite{black1976studies} and Christie \cite{christie1982stochastic}:
When an asset price declines, the associated company becomes automatically more leveraged since the ratio of its debt with respect to the equity value becomes larger. Hence the risk of the asset, namely its volatility, should become more important. Another economic interpretation of the leverage effect, inverting causality, is that the forecast of an increase of the volatility should be compensated by a higher rate of return, which can only be obtained through a decrease in the asset value, see \cite{campbell1992no,figlewski2000leverage,french1987expected}.\\ 

\noindent From an empirical viewpoint, leverage effect and the plausible interpretations for it have been widely studied in the literature, see for example \cite{bekaert2000asymmetric,bollerslev2006leverage,engle1993measuring,wu2002generalized}. Furthermore, some statistical methods enabling us to use high frequency data have been built to measure it, see \cite{ait2013leverage,wang2014estimation}. From a modeling perspective, the will to reproduce the leverage phenomenon has been a key motivation in the development of sophisticated time series models, for example of ARCH type, see \cite{bollerslev1992arch,ding1993long,nelson1991conditional,rodriguez2012revisiting,zakoian1994threshold}.
Finally, in financial engineering, it has become clear in the late eighties that it is necessary to introduce leverage effect in derivatives pricing frameworks in order to accurately reproduce the behavior of the implied volatility surface. This led to the rise of famous stochastic volatility models, where the Brownian motion driving the volatility is (negatively) correlated with that driving the price, see for example \cite{hagan2002managing,heston1993closed,hull1987pricing,stein1991stock} for SABR, Heston, Hull and White and Stein and Stein stochastic volatility models.\\

\noindent As mentioned above, traditional explanations for leverage effect are based on ``macroscopic" arguments from financial economics. In this paper, we wish to address the following question: Could microscopic interactions between agents
naturally lead to leverage effect at larger time scales? Hence we would like to know whether part of the foundations for leverage effect could be microstructural. To do so, our idea is to consider a very simple agent-based model, encoding well-documented and understood behaviors of market participants at the microscopic scale. Then we aim at showing that in the long run, this model leads to a price dynamic exhibiting leverage effect. This would demonstrate that typical strategies of market participants at the high frequency level naturally induce leverage effect.\\

\noindent One could argue that transactions take place at the finest frequencies and prices are revealed through order book type mechanisms. Therefore, it is an obvious fact that
leverage effect arises from high frequency properties. However, what we wish to show here is that under certain market conditions, typical high frequency behaviors,
having probably no connection with the financial economics concepts mentioned earlier, may give rise to some leverage effect at the low frequency scales.
It is important to emphasize that we do not claim that leverage effect should be fully explained by high frequency features. What we simply say is that part of it could be generated from the microstructure of the asset.\\

\noindent Another important stylized fact of financial data, which has been highlighted recently in \cite{gatheral2014volatility}, is the rough nature of the volatility process. Indeed, it is shown in \cite{gatheral2014volatility} that for a very wide range of assets, historical volatility time-series exhibit a behavior which is much rougher than that of a Brownian motion. More precisely, the dynamics of the log-volatility are typically very well modeled by a fractional Brownian motion with Hurst parameter around $0.1$, that is a process with H\"older regularity of order $0.1$. Furthermore, using a fractional Brownian motion with small Hurst index also enables us to reproduce very accurately the features of the volatility surface, see \cite{bayer2016pricing,gatheral2014volatility}.\\

\noindent The fact that for basically all reasonably liquid assets, volatility is rough, with the same order of magnitude for the roughness parameter, is of course very intriguing. Thus we also aim in this work at understanding how such a surprising feature can be generated. Some elements in this direction are already provided in \cite{jaisson2016rough}. Here we want to go further and investigate the behavior of the long term volatility in our microscopic model encoding the main stylized facts of modern market microstructure. We wish to show that the rough nature of the volatility naturally emerges from typical behaviors of market participants at the high frequency scale.\\

\noindent Our tick-by-tick price model is based on a bi-dimensional Hawkes process, very much inspired by the approaches in \cite{bacry2013modelling,bacry2013some,jaisson2015limit}. A bi-dimensional Hawkes process is a bivariate point process $(N_t^{+},N_t^{-})_{t\geq 0}$ taking values in $(\mathbb{R}_+)^2$ and with intensity $(\lambda_t^{+},\lambda_t^{-})$ of the form
$$\begin{pmatrix} \lambda_t^{+} \\ \lambda_t^{-} \end{pmatrix} = \begin{pmatrix} \mu^+ \\ \mu^- \end{pmatrix} +\int_0^t \begin{pmatrix} \varphi_1(t-s) & \varphi_3(t-s) \\ \varphi_2(t-s) & \varphi_4(t-s) \end{pmatrix} . \begin{pmatrix} dN_s^{+} \\ dN_s^{-} \end{pmatrix}.$$ Here $\mu^+$ and $\mu^-$ are positive constants and the functions $(\varphi_i)_{i=1,\ldots 4}$ are non-negative with associated matrix called kernel matrix, see Section \ref{parametrization} for further details. Hawkes processes have been introduced by Hawkes in \cite{hawkes1971point}. They are said to be self-exciting, in the sense that the instantaneous jump probability depends on the location of the past events. Hawkes processes are nowadays of standard use in finance, not only in the field of microstructure but also in risk management or contagion modeling, see among many others \cite{ait2015modeling,bacry2013modelling,bauwens2004dynamic,bowsher2007modelling,chavez2005point,embrechts2011multivariate,errais2010affine,jaisson2015limit,jaisson2016rough}. It is explained in \cite{bacry2013modelling} that a relevant model for the ultra high frequency dynamic of the price $P_t$ of a large tick asset\footnote{A large tick asset is an asset whose bid-ask spread is almost always equal to one tick and therefore essentially moves by one tick jumps, see \cite{dayri2012large}.} is simply given by
$$P_t=N_t^{+}-N_t^{-}.$$
Thus, in this approach, $N_t^{+}$ corresponds to the number of upward jumps of the asset in the time interval $[0,t]$ and $N_t^{-}$ to the number of downward jumps. Hence, the instantaneous probability to get an upward (downward) jump depends on the arrival times of the past upward and downward jumps. Furthermore, by construction, the price process lives on a discrete grid, which is obviously a crucial feature of high frequency prices in practice. Statistical properties of this model have been studied in details in \cite{bacry2013modelling}. In particular, it is shown that such dynamic is very convenient in order to reproduce the  commonly observed bid-ask bounce effect.\\

\noindent This simple tick-by-tick price model enables us to encode very easily the following important stylized facts of modern electronic markets in the context of high frequency trading:
\begin{itemize}
\item[$i)$] Markets are highly endogenous, meaning that most of the orders have no real economic motivation but are rather sent by algorithms in reaction to other orders, see \cite{filimonov2015apparent,hardiman2013critical} and Section \ref{sec_endo} for more details.
\item[$ii)$] Mechanisms preventing statistical arbitrages take place on high frequency markets. Indeed, at the high frequency scale, building strategies which are on average profitable is hardly possible, see \cite{abergel2014understanding}.
\item[$iii)$] There is some asymmetry in the liquidity on the bid and ask sides of the order book. This simply means that buying and selling are not symmetric actions. Indeed, consider for example a market maker, with an inventory which is typically positive. He is likely to raise the price by less following a buy order than to lower the price following the same size sell order. This is because its inventory becomes smaller after a buy order, which is a good thing for him, whereas it increases after a sell order, see \cite{brennan2012sell,brunnermeier2009market,hendershott2006market,ho1981optimal,tayal2012measuring}. 
\item[$iv)$] A significant proportion of transactions is due to large orders, called metaorders, which are not executed at once but split in time by trading algorithms, see \cite{almgren2001optimal,lehalle2013market}.
\end{itemize}
\noindent In a Hawkes process framework, the first of these properties corresponds to the case of so-called {\it nearly unstable Hawkes processes}, that is Hawkes processes for which the stability condition is almost saturated. This means the spectral radius of the kernel matrix integral is smaller than but close to unity, see \cite{filimonov2015apparent,hardiman2013critical,jaisson2015limit,jaisson2016rough}. The second and third ones impose a specific structure on the kernel matrix and the fourth one leads to functions $\varphi_i$ with heavy tails, see \cite{jaisson2016rough}. The parametrization of our price process corresponding to the four properties above is developed in more details in Sections \ref{parametrization} and \ref{parametrization2}.\\

\noindent In this work, we study the long term behavior of such Hawkes-based ultra high frequency price models, for which the parameters are consistent with the four mentioned properties of market microstructure. Doing so, we
investigate the macroscopic price dynamics arising from a situation where the four ingredients above are put together. More precisely, we start with the case of a Hawkes-based model where
Properties $i$, $ii$ and $iii$ only are satisfied. Our first result states that in this setting, the macroscopic dynamic of the price is that of a Heston stochastic volatility model as introduced in \cite{heston1993closed}, where the volatility
is (negatively) correlated with the price. Hence leverage effect is produced. This extends some results in \cite{jaisson2015limit} where a non-correlated Heston limit is obtained.
Then, when in addition Property $iv$ is encoded in our microscopic model, we show that a so-called rough-Heston model, where the volatility is rough and negatively correlated with the price, is generated at low frequency. More precisely, as in \cite{gatheral2014volatility}, the volatility process is driven by a fractional Brownian motion with Hurst parameter smaller than 1/2.\\

\noindent Of course our results are not the first ones relating high frequency dynamics to long term behaviors with stochastic volatility.
The most famous example is probably that of Nelson who shows in \cite{nelson1990arch} that in specific settings, GARCH processes converge to
(uncorrelated) stochastic volatility models, see also \cite{corradi2000reconsidering,duan1997augmented,lindner2009continuous}. However, to our knowledge,
we provide the first natural, non ad-hoc approach allowing for leverage effect, and even rough volatility, in the long term limit of the price dynamic.\\

\noindent The paper is organized as follows. In Section \ref{Part1}, we parametrize our Hawkes-based microscopic price model so that Properties $i$, $ii$ and $iii$ are satisfied. Then we show that after proper rescaling, this price converges in the long run to a Heston stochastic volatility model where leverage effect is observed. In Section \ref{Part2}, we incorporate Property $iv$ into our microscopic model and prove that it leads to a rough Heston model at the macroscopic scale, where leverage effect is still generated. Some proofs are relegated to Section \ref{proofs} and some useful technical results are given in an appendix.

\section{{From high frequency features to leverage effect}}\label{Part1}
We build in this section a Hawkes-based microscopic tick-by-tick model in which Properties $i$, $ii$ and $iii$ are satisfied. This leads us to a specific parametrization of our Hawkes process. We show that after suitable rescaling, the long term price dynamic becomes that of a Heston model. We start by defining our microscopic price model.

\subsection{Building a suitable microscopic price model}
\label{parametrization}
\subsubsection{The Hawkes process framework}
We consider a tick-by-tick price model based on a bi-dimensional Hawkes process $N_t = (N_t^+,N_t^-)$, with intensity $\lambda_t = (\lambda_t^+,\lambda_t^-)$ defined by 
$$\begin{pmatrix} \lambda_t^{+} \\ \lambda_t^{-} \end{pmatrix} = \begin{pmatrix} \mu^+ \\ \mu^- \end{pmatrix} +\int_0^t \begin{pmatrix} \varphi_1(t-s) & \varphi_3(t-s) \\ \varphi_2(t-s) & \varphi_4(t-s) \end{pmatrix} . \begin{pmatrix} dN_s^{+} \\ dN_s^{-} \end{pmatrix},$$ 
where $\mu^+$ and $\mu^-$ are positive constants and 
$$\phi= \begin{pmatrix} \varphi_1 & \varphi_3 \\ \varphi_2 & \varphi_4\end{pmatrix}: \mathbb R_+ \rightarrow {\cal{M}}^2(\mathbb R_+^*)$$
is a kernel matrix whose components $\varphi_i$ are positive and locally integrable. Inspired by \cite{bacry2013modelling,bacry2013some,jaisson2015limit}, our model for the ultra high frequency transaction price $P_t$ is simply given by
$$P_t = N_t^+ - N_t^-.$$
Thus $N_t^+$ is the number of upward jumps of one tick of the asset in the time interval $[0,t]$ and $N_t^-$ is the number of downward jumps of one tick of the asset in the time interval $[0,t]$.\\

\noindent Let us now interpret the intensity process $\lambda_t^+$ (interpretation for $\lambda_t^-$ goes similarly). At time $t$, the probability to get a new one-tick upward jump between $t$ and $t+dt$ is given by $\lambda_t^+ dt$. This probability can be decomposed into three terms:
\begin{itemize}
\item $\mu_+ dt$, which is the Poissonian part of the intensity and therefore corresponds to the probability that the price goes up because of some exogenous reason.
\item $\big(\displaystyle\int_0^t \varphi_1(t-s) dN_s^+\big)dt$, which is the probability of upward jump induced by past upward jumps.
\item $\big(\displaystyle\int_0^t \varphi_3(t-s) dN_s^-\big)dt$, which is the probability of upward jump induced by past downward jumps.
\end{itemize}
\noindent In particular, we see here that when the $\varphi_i$ have suitable shapes, it is easy to reproduce the bid-ask bounce effect by imposing a high probability of upward (resp. downward) jump right after a downward (resp. upward) jump. 
\subsubsection{Encoding Properties $ii$ and $iii$}

\noindent We now provide a specific structure on the parameters of the intensity process so that Properties $ii$ and $iii$ are satisfied in our model. Property $ii$ is the no-statistical arbitrage condition. In a high frequency setting, this amounts to say that on average, there should be essentially as many upward as downward jumps on any given time-period. We translate this within our Hawkes framework noting that
$$\E[N_t^+]=\int_{0}^t\E[\lambda_s^+] ds,~~\E[N_t^-]=\int_{0}^t\E[\lambda_s^-] ds,$$
and 
\begin{align*}
\E[\lambda_t^+]&=\mu^++\int_0^{t}\!\varphi_1(t-s)\E[\lambda_s^+]ds+\int_0^{t}\!\varphi_3(t-s)\E[\lambda_s^-]ds,\\\E[\lambda_t^-]&=\mu^-+\int_0^{t}\!\varphi_2(t-s)\E[\lambda_s^+]ds+\int_0^{t}\!\varphi_4(t-s)\E[\lambda_s^-]ds.
\end{align*}
Therefore we obtain that a simple and natural way to implement the no-statistical arbitrage condition is to set $\E[\lambda_t^+] = \E[\lambda_t^-] $ by imposing 
$$ \mu^+ = \mu^-~~\text{ and }~~\varphi_1 + \varphi_3 = \varphi_2 + \varphi_4.$$

\noindent In term of microscopic price movements, Property $iii$, which states that the ask side is more liquid than the bid side, can be translated as follows: the conditional probability to observe an upward jump right after an upward jump is smaller than the conditional probability to observe a downward jump right after a downward jump. In our Hawkes framework, it amounts to have $\varphi_1(x)<\varphi_4(x)$ or similarly $\varphi_3(x)>\varphi_2(x)$ when $x$ is close to zero. For simplicity and technical convenience, we in fact make the more restrictive assumption that there exists some $\beta>1$ such that 
$$\varphi_3=\beta\varphi_2.$$

\noindent Therefore we assume the following structure for the intensity process:
\begin{equation}\label{intensityStructure}
\begin{pmatrix} \lambda_t^{+} \\ \lambda_t^{-} \end{pmatrix} = \mu \begin{pmatrix} 1 \\ 1 \end{pmatrix} +\int_0^t \phi(t-s). \begin{pmatrix} dN_s^{+} \\ dN_s^{-} \end{pmatrix},
\end{equation}
where 
$$\phi= \begin{pmatrix} \varphi_1 & \beta \varphi_2 \\ \varphi_2 & \varphi_1 + (\beta-1)\varphi_2 \end{pmatrix},$$
with $\mu>0$ and $ \beta\geq 1$. We now explain how to deal with Property $i$.

\subsubsection{Dealing with Property $i$: Nearly unstable Hawkes processes}\label{sec_endo}

Property $i$ states that modern markets are highly endogenous. To understand how this high degree of endogeneity can be translated through our Hawkes-based price model, let us consider for simplicity a one-dimensional Hawkes process $\tilde{N}_t$ with intensity 
$$ \tilde{\lambda}_t = \tilde{\mu} + \int_0^t \tilde{\varphi}(t-s) d\tilde{N}_s,$$
where $\tilde{\mu}>0$ and $\tilde{\varphi}$ is a non-negative measurable function such that its $L^1$ norm $||\tilde{\varphi}||_1$ satisfies $||\tilde{\varphi}||_1<1$. This last constraint is called stability condition and plays the same role as that which states that the coefficient of an order $1$ auto-regressive process has to be smaller than one, see \cite{jaisson2015limit}. In particular, this condition ensures the existence of a stationary solution for the intensity (when time starts at $-\infty$). Such one-dimensional Hawkes processes are usually considered to model order flows, see \cite{bacry2015hawkes} and the references therein. So $\tilde{N}_t$ can typically be viewed as the number of transactions in the time interval $[0,t]$.\\

\noindent From a probabilistic viewpoint, the cluster representation of Hawkes processes, see \cite{hawkes1974cluster}, enables to see $\tilde{N}$ as a population process. In this population, migrants arrive following a Poisson process with intensity $\tilde{\mu}$. Each migrant gives birth to children according to an inhomogeneous Poisson process with intensity $\tilde{\varphi}$. Then each child also gives birth to children according to an inhomogeneous Poisson process with intensity $\tilde{\varphi}$ and so on. Coming back to financial markets, let us consider a dichotomy between ``economic" (or exogenous) orders, which are executed because some market participants have a fundamental will to buy or sell, and endogenous orders, which are just sent in reaction to other orders. In the Hawkes context, it is therefore very natural to make the following interpretation: exogenous orders correspond to migrants and endogenous orders to descendants of migrants, see \cite{filimonov2015apparent,hardiman2013critical,jaisson2015limit}.\\

\noindent Now remark that each migrant, or descendant of a migrant, has on average $\|\tilde{\varphi}\|_1$ children. Hence a migrant has on average
$$ \sum_{k \geq 1} \|\tilde{\varphi}\|_1^k = \frac{\|\tilde{\varphi}\|_1}{1-\|\tilde{\varphi}\|_1}$$
descendants. Now, the number of people in a ``family" being the number of descendants plus one (the plus one corresponding to the initial migrant), the proportion of descendants in the whole population is given by $||\tilde{\varphi}||_1$. In our financial interpretation, it means that $||\tilde{\varphi}||_1$ corresponds to the proportion of endogenous orders in the market. Hence, to get a model which is in agreement with Property $i$, we need to take $||\tilde{\varphi}||_1$ smaller than but close to unity. This situation is called nearly unstable case, and is actually in agreement with the empirical measurements for $||\tilde{\varphi}||_1$ made in \cite{bacry2012non,filimonov2015apparent,hardiman2013critical}.\\

\noindent Let us now come back to our bi-dimensional Hawkes process of interest with intensity defined by \eqref{intensityStructure}. In the same way as in the one-dimensional case, one can define the degree of endogeneity as the spectral radius of the kernel matrix integral, that is 
$$ {\cal S} (\int_0^\infty \phi(s)ds) = \|\varphi_1\|_1 + \beta \|\varphi_2\|_1,$$where ${\cal S}$ denotes the spectral radius operator. We want to assume that this spectral radius is smaller than but close to unity. To do so, we introduce an asymptotic framework, in the spirit of \cite{jaisson2015limit,jaisson2016rough}.
More precisely, we work on a sequence of probability spaces $({\Omega}^T, {\cal{F}}^T,{\mathbb{P}}^T)$ , indexed by $T>0$, on which $N^T = (N^{T,+},N^{T,-})$ is a bi-dimensional Hawkes process defined on $[0,T]$ and with intensity of the form
\begin{equation} 
\label{intensity}
\lambda_t^T=  \begin{pmatrix}\lambda_t^{T,+} \\  \lambda_t^{T,-}  \end{pmatrix} = {\mu}_T \begin{pmatrix} 1  \\ 1   \end{pmatrix}   + \int_0^t \phi^T(t-s). dN_s^T.
\end{equation}
For given $T$, the probability space is equipped with the filtration $(\mathcal{F}_t^T)_{t\geq 0}$, where $\mathcal{F}_t^T$ is the $\sigma$-algebra generated by 
$(N_s^T)_{s \leq t}$. Respecting the constraints on the parameters given in \eqref{intensityStructure} and taking into account the discussion above about the endogeneity of the market, we make the following assumption on $\lambda_t^T$.

\begin{assumption}\label{assumptionStructure} 
We have $\mu_T>0$ and
$$ \phi^T = a_T \phi,~~\phi = \begin{pmatrix} \varphi_1 & \beta \varphi_2  \\ \varphi_2 & \varphi_1 + (\beta-1) \varphi_2  \end{pmatrix},$$
where $\beta\geq 1$, $\varphi_1$ and $\varphi_2$ are two positive measurable functions such that 
$$ {\cal S} (\int_0^\infty \phi(s)ds) = \|\varphi_1\|_1 + \beta \|\varphi_2\|_1 = 1 $$
and $a_T$ is an increasing sequence of positive numbers converging to one.
\end{assumption}
\noindent From now on, our microscopic price process is given by
$$P_t^T=N^{T,+}-N^{T,-}.$$
Thus, under Assumption \ref{assumptionStructure}, we are indeed working in the nearly unstable case since $${\cal S} (\int_0^\infty \phi^T(s)ds)=a_T.$$
Therefore, our microscopic price process $P^T$ reproduces Properties $i$, $ii$ and $iii$. We now focus on the asymptotic behavior of $P^T$.
 
\subsection{The macroscopic limit with leverage of the high frequency model}
We give in this section our convergence result for the microscopic price towards a Heston model. In fact such result can be found in \cite{jaisson2015limit} in the case $\beta = 1$. As in \cite{jaisson2015limit}, we need to consider the following assumption on the asymptotic framework and the kernel function.
\begin{assumption}\label{smallTail} 
There exist positive parameters $\lambda$, $\mu$ and $m$ such that
$$ T(1-a_T) \underset{T \rightarrow \infty}{\rightarrow}  \lambda, \quad \mu_T = \mu,$$
and 
$${\cal S} ( \int_0^\infty x \phi(x)  dx )=m<\infty.$$
\end{assumption}
\noindent It is well-explained in \cite{jaisson2015limit} that assuming that the kernel $L^1$ norm $a_T$ goes to unity in such a way that $T(1-a_T)$ is of order one is the only asymptotic framework enabling us to recover a non-degenerate limit. Now let $$ \psi^T= \sum_{k\geq 1}  (\phi^T)^{*k},$$
where  $(\phi^T)^{*1} = \phi^T $ and for $k>1$, $(\phi^T)^{*k}(t) = \int_0^t \phi^T(s)(\phi^T)^{*(k-1)}(t-s)ds $. The following technical assumption is also required in \cite{jaisson2015limit}.
\begin{assumption}\label{regularity} 
The function $\psi^T$ is uniformly bounded and $\phi$ is differentiable such that each component $\phi_{ij}$ satisfies $||\phi_{ij}'||_{\infty} < \infty$ and $||\phi_{ij}'||_{1} < \infty$.
\end{assumption}
\noindent The uniform boundedness assumption here is not really restrictive. Indeed, as it will be clear from the computations in Section \ref{intuition1}, a sufficient condition for it is the fact that the largest eigenvalue of $\phi$ is non-increasing, see also \cite{jaisson2015limit}. In our model, this is for example the case if both $\varphi_1$ and $\varphi_2$ are non-increasing.\\

\noindent When $\beta = 1$, under Assumptions \ref{assumptionStructure}, \ref{smallTail} and \ref{regularity}, it is proved in \cite{jaisson2015limit} that the rescaled price process
$$ \frac{1}{T}P_{tT}^T = \frac{N_{tT}^{T,+}- N_{tT}^{T,-}}{T}$$
converges in law over $[0,1]$ towards a Heston model defined by 
$$ P_t= \frac{1}{1-(\|\varphi_1\|_1 -\|\varphi_2\|_1)}\int_0^t \sqrt{X_s} dW_s,$$
with
$$ dX_t = \frac{\lambda}{m} (\frac{2 \mu}{\lambda}- X_t) dt + \frac{1}{m} \sqrt{X_t} dB_t,~~X_0 = 0,$$
where $W$ and $B$ are two independent Brownian motions.\\

\noindent However, when $\beta = 1$, the important Property $iii$ about the liquidity asymmetry between the bid and ask sides of the order book is not reproduced in the dynamic of the microscopic price. Our first main theorem below shows that this property, encoded by the fact that $\beta>1$, is the microscopic feature at the origin of leverage effect at low frequency.

\begin{theorem} \label{Heston1} 
Under Assumptions \ref{assumptionStructure}, \ref{smallTail} and \ref{regularity}, as $T$ tends to infinity, the rescaled microscopic price 
$$ \frac{1}{T} P_{tT}^T = \frac{N_{tT}^{T,+}-N_{tT}^{T,-}}{T},~~t \in [0,1],$$
converges in law for the Skorokhod topology to the following Heston model:
$$ P_t = \frac{1}{1-(\|\varphi_1\|_1 -\|\varphi_2\|_1)}\sqrt{\frac{2}{1+\beta}} \int_0^t \sqrt{X_s} dW_s,$$
with
$$ dX_t = \frac{\lambda}{m} \big((\beta +1) \frac{\mu}{\lambda} - X_t\big) dt + \frac{1}{m} \sqrt{ \frac{1+\beta^2}{1+\beta}} \sqrt{X_t} dB_t,~~X_0 =0,$$
where $(W,B)$ is a correlated bi-dimensional Brownian motion with $$d\langle W,B \rangle_t= \frac{1-\beta}{\sqrt{2(1+\beta^2)}}dt.$$
\end{theorem}

\noindent Hence, putting Properties $i$, $ii$ and $iii$ together in a simple but reasonable way (through the microscopic price $P^T$), we naturally obtain stochastic volatility and leverage effect in the long run. Indeed, when $\beta>1$, the asymmetry in the liquidity at the microstructural level generates a negative correlation between low frequency price returns and volatility increments. Nevertheless, Properties $i$ and $ii$ are also crucial in order to obtain Theorem \ref{Heston1}. In fact, no stochastic volatility can be obtained without Property $i$ and the failure of Property $ii$ would lead to a drift process in the limit.\\

\noindent Finally, note that from a technical point of view, to our knowledge, this result is the first scaling limit of a microscopic price process inducing leverage effect in the long run in a non ad-hoc way.\\

\noindent We now give in the next section a general result about the convergence of nearly unstable multidimensional Hawkes processes. This result is the key element of the proof of Theorem \ref{Heston1}.

\subsection{Convergence of nearly unstable multidimensional Hawkes processes}
\label{dimensional}
\subsubsection{Setting} \label{setting}
\noindent In order to show Theorem \ref{Heston1}, we study the convergence of a general sequence of nearly unstable $d$-dimensional Hawkes processes defined on $[0,T]$, with $T$ tending to infinity. We keep the notation $N^T$ for our Hawkes process of interest whose intensity $ \lambda^T$ is defined by
$$ \lambda_t^T = \mu_T \mathbf{1} + \int_0^t \phi^T(t-s).dN_s^T, $$
where $\mu_T>0$ and $\phi^T = a_T \phi$, with $a_T$ an increasing sequence of positive numbers converging to unity, and the matrix $\phi: \mathbb R_+ \rightarrow {\cal{M}}^d (\mathbb R_+^*)$ has integrable components such that
$${\cal S}(\int_0^\infty\phi(s)ds) = 1.$$
We furthermore assume that for any $t\geq 0$, $\phi(t)$ is diagonalizable on $\mathbb R$. We write $\lambda_1(t) \geq .. \geq \lambda_d(t)$ for the eigenvalues of $\phi(t)^*$ (here $^*$ refers to the transpose operator) and $v_1,...,v_d$ for the corresponding eigenvectors. We assume that these eigenvectors do not depend on $t$ (as it is the case under Assumption \ref{assumptionStructure}). We also recall that from
Frobenius-Perron theorem, for $i\geq 2$, $|\lambda_i(t)|<\lambda_1(t) = {\cal S}\big(\phi(t)\big)$ and $v_1$ can be taken in $\mathbb{R}_+^d$.

\subsubsection{Intuition for the result and theorem}\label{intuition1}

We now provide some non-rigorous developments which are helpful in order to understand the asymptotic behavior of the multidimensional process $N^T$. We work under Assumptions \ref{smallTail} and \ref{regularity}. Let
$$ M_t^T= N_t^T - \int_0^t \lambda_s^T ds$$
be the martingale associated to $N^T$. We have
$$\lambda_t^T = \mu_T \textbf{1} + \int_0^t \phi^T(t-s).dM_s^T + \int_0^t \phi^T(t-s).\lambda_s^T ds.$$
Using Lemma \ref{hopf} in Appendix together with Fubini theorem and the fact that the convolution product $\psi^T*\phi^T$ satisfies $\psi^T*\phi^T=\psi^T-\phi^T$, we get
\begin{equation}\label{intensity3}
\lambda_t^T = \mu_T \mathbf{1} +\mu_T \int_0^t \psi^T(t-s)ds.\mathbf{1}+ \int_0^t \psi^T(t-s).dM_s^T,
\end{equation}
where $$\psi^T = \sum_{k \geq 1} (\phi^T)^{*k} = \sum_{k \geq 1} a_T^k \phi^{*k}.$$ Therefore
\begin{equation} \label{expectlambda} 
\mathbb{E}[\lambda_t^T]= \mu_T \mathbf{1} +\mu_T \int_0^t \psi^T(t-s)ds.\mathbf{1}
\end{equation}
and 
$$ \mathbb{E}[\lambda_{tT}^T] = \mu_T \mathbf{1} +\mu_T T \int_0^t \psi^T\big(T(t-s)\big)ds.\mathbf{1}.$$
The function $\psi^T$ being uniformly bounded and $\mu_T$ being constant equal to $\mu >0$, we get that $\lambda_{tT}^T$ is of order $T$. Thus a natural rescaling in time and space leads us to consider for $t \in [0,1]$ the process
$$C_t^T= \frac{1}{T} \lambda_{tT}^T.$$
From \eqref{intensity3}, we obtain
$$C_t^T = \frac{\mu}{T} \mathbf{1} + \mu \int_0^t \psi^T\big(T(t-s)\big) ds . \mathbf{1} + \int_0^t 
 \psi^T\big(T(t-s)\big) . d\overline{M}_s^T ,$$
with $\overline{M}_t^T =M_{tT}^T/T$. Note that since $$\langle M^T,M^T\rangle_t = diag(\int_0^t \lambda_s^T ds),$$ we get that
$$
{\mathbb{E}}[\langle \overline{M}^T ,\overline{M}^T \rangle_t]=\frac{1}{T^2}{\mathbb{E}}\big[ diag(\int_0^{tT} \lambda_s^T ds)\big] = diag(\int_0^{t} {\mathbb{E}}[C_s^T] ds)
$$ is bounded. Now remark that for each $i \in \{1,...,d\}$, using a recursion, we easily see that for any $k \geq 1$, $v_i^*.\phi^{*k}(t) = \lambda_i^{*k}(t) v_i^*$. Consequently, defining for $i \in \{1,...,d\}$
$$\psi_i^T= \sum_{k \geq 1} a_T^k \lambda_i^{*k},$$
we have
$$v_i^*.\psi^T = \psi_i^T v_i^*.$$ 
Hence we can write the dynamic of $v_i^*. C_t^T$ as follows:
\begin{equation} \label{coordonateRescaledIntensity}
v_i^*. C_t^T = \frac{\mu}{T} (v_i^*.\mathbf{1}) + \mu (v_i^*. \mathbf{1}) \int_0^t \psi_i^T\big(T(t-s)\big) ds  + \int_0^t  \psi_i^T\big(T(t-s)\big) (v_i^*.d\overline{M}_s^T).
 \end{equation}

\noindent Thus, to understand the asymptotic behavior of $v_i^*.C^T$ as $T$ goes to infinity, we need to study that of the functions $\psi_i^T(T.)$. To do so,
one can compute the Fourier transform $\hat{\psi}^T_j(T.)$ of $\psi_j^T(T.)$ for each $j \in \{1,...,d\}$. We have
$$
\hat{\psi}^T_j(T.)(z)= \int_{x \in \mathbb R_+} \psi^T_j(Tx) e^{ixz} dx = \frac{1}{T} \sum_{k \geq 1} a_T^k  \big(\hat{\lambda}_j(z/T)\big)^k = \frac{ a_T  \hat{\lambda}_j(z/T)}{T\big(1- a_T  \hat{\lambda}_j(z/T)\big)}.$$
Now, as $T$ goes to infinity, $\hat{\lambda}_j(z/T)$ tends to $\| \lambda_j \|_1$ and recall that $\| \lambda_j \|_1 < 1 $ for $j \geq 2 $. Thus, for $j \geq 2 $, $\psi_j^T(T.)$ should asymptotically vanish, as should consequently be the case for $v_j^*.C^T$.\\

\noindent For $j=1$, using Assumption \ref{smallTail} we have
$$\underset{T \rightarrow \infty}{\lim}T (  \hat{\lambda}_1(z/T) - 1) = i z \int_0^\infty x \lambda_1(x) dx = izm.$$ Therefore, in that case,
$$\hat{\psi}^T_1(T.)(z) = \frac{ a_T  \hat{\lambda}_1(z/T)}{T(1- a_T) - a_T T (  \hat{\lambda}_1(z/T) - 1)} \underset{T \rightarrow \infty}{\rightarrow} \frac{1}{\lambda - izm},$$
which is the Fourier transform of
$$ x \in \mathbb R_+ \rightarrow \frac{1}{m} e^{-\frac{\lambda}{m}x}.$$
Hence we can expect that $\psi_1(Tx)$ converges to $\frac{1}{m} e^{-\frac{\lambda}{m}x}$.\\

\noindent Let us now deduce from the preceding computation the behavior of $v_1^*.C^T$. From \eqref{coordonateRescaledIntensity}, this quantity can be written
\begin{equation}\label{v1RescaledIntensity}
v_1^*. C_t^T = \frac{\mu}{T} (v_1^*.\mathbf{1}) + \mu (v_1^*. \mathbf{1})\int_0^t \psi_1^T\big(T(t-s)\big) ds + \int_0^t  \psi_1^T\big(T(t-s)\big) \sqrt{(v_1^2)^*.C_s^T} dB_s^T,
\end{equation}
where $v_1^2 = (v_{1,i}^2)_{1\leq i \leq d}$ and 
\begin{equation}\label{brownian1}
B_t^T= \int_0^{tT} \frac{v_1^*.dM_s^T}{\sqrt{T (v_1^2)^*.\lambda_s^T}}.
\end{equation}
The sequence of processes $B^T$ has been specifically chosen since the associated sequence of quadratic variations converges to identity. Thus the limit of $B^T$ is a Brownian motion.\\

\noindent Now define an orthonormal basis $(e_1,..,e_d)$ of $\mathbb{R}^d$ such that $e_1^*.v_1>0$ and
$$ span(e_2,..,e_d) = span(v_2,..,v_d)$$
and set $v'= e_1 - \frac{1}{e_1^*.v_1} v_1$. Note that $v'$ belongs to $span(v_2,..,v_d)$.
Decomposing $v_1^2$ in the basis $(e_1,...,e_d)$, we get
$$ (v_1^2)^*.C_t^T = \frac{e_1^*.v_1^2}{e_1^*.v_1}(v_1^*.C_t^T) + (e_1^*.v_1^2) \big((v')^*.C_t^T\big) + \sum_{2 \leq i \leq d} (e_i^*.v_1^2) (e_i^*.C_t^T).$$
Thus, since for any vector $v \in span(v_2,...,v_d)$, $v^*.C_t^T$ converges to zero, we deduce that $(v_1^2)^*.C_t^T$ has the same asymptotic behavior as
$$ \frac{e_1^*.v_1^2}{e_1^*.v_1}(v_1^*.C_t^T).$$
Therefore, letting $T$ go to infinity in \eqref{v1RescaledIntensity}, we can expect $v_1^*.C_t^T$ to be solution of the following stochastic differential equation:
$$ X_t =  \frac{\mu}{m} \int_0^t e^{-\frac{\lambda}{m}(t-s)} ds (v_1^*.\mathbf{1}) +\frac{1}{m} \sqrt{\frac{e_1^*.v_1^2}{e_1^*.v_1}} \int_0^t  e^{-\frac{\lambda}{m}(t-s)} \sqrt{X_s}dB_s.$$
This exactly corresponds to a Cox-Ingersoll-Ross process since it can be rewritten
$$ dX_t = \frac{\lambda}{m}\big(\frac{\mu}{\lambda}(v_1^*.\mathbf{1}) - X_t\big) dt + \frac{1}{m} \sqrt{\frac{e_1^*.v_1^2}{e_1^*.v_1}} \sqrt{X_t} dB_t,~~X_0 = 0.$$
Hence, using the decomposition of $C_t^T$ in the basis $(e_1,...,e_d)$ given by
$$ C_t^T = \frac{1}{e_1^*.v_1} (v_1^*.C_t^T) e_1+ \big((v')^*.C_t^T\big) e_1+ \sum_{2 \leq i \leq d} (e_i^*.C_t^T) e_i,$$
we finally obtain the following theorem whose rigorous proof is given in Section \ref{proof1}.
\begin{theorem} 
\label{Convergence1}
Under the setting of Section \ref{setting} together with Assumptions \ref{smallTail} and \ref{regularity}, the multidimensional process $(C_t^T, B_t^{T})_{t \in [0,1]}$ converges in law for the Skorokhod topology to $(\frac{1}{e_1^*.v_1} X e_1, B)$ where $B$ is a Brownian motion and $X$ satisfies the following (one-dimensional) Cox-Ingersoll-Ross dynamic:
$$ dX_t = \frac{\lambda}{m}\big(\frac{\mu}{\lambda}(v_1^*.\mathbf{1}) - X_t\big) dt + \frac{1}{m} \sqrt{\frac{e_1^*.v_1^2}{e_1^*.v_1}} \sqrt{X_t} dB_t,~~ X_0= 0.$$
\end{theorem}

\noindent Theorem \ref{Convergence1} is a general result about the asymptotic behavior of multidimensional nearly unstable Hawkes processes. We see in particular that the non-degeneracy concentrates around the first eigenvector. Also, from Theorem \ref{Convergence1}, we obtain an immediate corollary given below which will enables us to prove Theorem \ref{Heston1}.

\subsubsection{Application to our microscopic model}

Let us consider a bi-dimensional Hawkes processes sequence $N^T = (N^{T,+},N^{T,-})$ with intensity $\lambda^T = (\lambda^{T,+},\lambda^{T,-})$ as in Assumption \ref{assumptionStructure}. In this case, 
$$\lambda_1 = \varphi_1+ \beta \varphi_2,~~\lambda_2 = \varphi_1-\varphi_2,$$ 
and  
$$v_1 = \begin{pmatrix} 1 \\ \beta\end{pmatrix},~~v_2 = \begin{pmatrix} 1 \\ -1\end{pmatrix}.$$ 
We therefore have the following corollary of Theorem \ref{Convergence1} which will lead us to the long term limit of our microscopic price model.
\begin{corollary} 
\label{corol1}
Under Assumptions \ref{assumptionStructure}, \ref{smallTail} and \ref{regularity}, the process $(C_t^{T,+}, C_t^{T,-}, B_t^{T})_{t \in [0,1]}$ converges in law for the Skorokhod topology to $(\frac{1}{\beta+1} X,\frac{1}{\beta+1} X, B)$ where $B$ is a Brownian motion and $X$ satisfies the following (one-dimensional) Cox-Ingersoll-Ross dynamic:
$$ dX_t = \frac{\lambda}{m}(\frac{\mu}{\lambda}(\beta+1) - X_t) dt + \frac{1}{m} \sqrt{\frac{1+\beta^2}{1+\beta}} \sqrt{X_t} dB_t,~~X_0 = 0.$$
\end{corollary}

\noindent Here $X$ essentially corresponds to a limiting volatility process. The Brownian motion in the dynamic of $X$ comes from the limit of $B^T$, the process defined in \eqref{brownian1} and driven by $v_1^*.dM_s^T$. In our microscopic model, $M^T=(M^{T,+},M^{T,+})$. As will be clear from the proof of Theorem \ref{Heston1}, the Brownian motion driving the price in Theorem \ref{Heston1} arises from the limiting behavior of $M^{T,+}-M^{T,+}$. Hence, the emergence of leverage effect in the limit is due to the non-zero covariation between $v_1^*.dM_s^T$ and $M^{T,+}-M^{T,+}$.

\section{From high frequency features to rough volatility}\label{Part2}

\subsection{Encoding Property $iv$}\label{parametrization2}

In Section \ref{Part1}, we have built a microscopic Hawkes-based price model compatible with Properties $i$, $ii$ and $iii$. Theorem \ref{Heston1} states that it converges in the long run to a classical Heston model. However, Property $iv$, that is the wide presence of metaorders on the market, which is a crucial feature of high frequency markets, is not encoded in such model. As explained in \cite{jaisson2016rough}, this can be translated in the Hawkes framework by considering the model defined by Assumption \ref{assumptionStructure} but under the condition that the kernel matrix exhibits a heavy tail, as observed in practice, see \cite{bacry2014estimation,hardiman2013critical}. Consequently, we need to replace Assumption \ref{smallTail} in order to get a slowly decreasing behavior for the kernel matrix. This also implies a modification of the asymptotic setting in order to retrieve a non-degenerate scaling limit, see \cite{jaisson2016rough}. More precisely, in this section, instead of Assumption \ref{smallTail} we consider the following one:
\begin{assumption}\label{fatTail}
There exist $\alpha\in(1/2,1)$ and $C>0$ such that 
$$ \alpha x^\alpha \int_x^\infty \lambda_1(s) ds \underset{x \rightarrow \infty}{\rightarrow} C.$$
Moreover, for some $\lambda^* >0$ and $\mu>0 $,
$$ T^\alpha (1-a_T) \underset{T \rightarrow \infty}{\rightarrow}  \lambda^* >0,~~T^{1-\alpha}\mu_T \underset{T \rightarrow \infty}{\rightarrow} \mu.$$
\end{assumption}
\noindent Of course, the first eigenvalue under Assumption \ref{assumptionStructure} being $\varphi_1+\beta\varphi_2$, Assumption \ref{fatTail} on $\lambda_1$ can also be expressed in term of the asymptotic behavior of $\varphi_1$ and $\varphi_2$. Note that in practice, estimated values for $\alpha$ are actually close to $1/2$, see \cite{bacry2014estimation,hardiman2013critical}. We now give the asymptotic behavior of our price model under Assumption \ref{fatTail}.

\subsection{The rough macroscopic limit of the high frequency model}

Let $\lambda=\alpha \lambda^*/\big(C \Gamma(1-\alpha)\big)$. We have the following result for the long term limit of our microscopic model compatible with Properties $i$, $ii$, $iii$ and $iv$.

\begin{theorem}\label{finalHeston}
Under Assumptions \ref{assumptionStructure} and \ref{fatTail}, as $T$ tends to infinity, the rescaled microscopic price
$$\sqrt{\frac{1-a_T}{\mu T ^\alpha}} P_{tT}^T,~~t \in [0,1],$$ converges in the sense of finite dimensional laws to the following rough Heston model:
$$ P_t = \frac{1}{1-(\|\varphi_1\|_1 -\|\varphi_2\|_1)}\sqrt{\frac{2}{\beta+1}} \int_0^t \sqrt{Y_s} dW_s,$$
with $Y$ the unique solution of 
$$ Y_t = \frac{1}{\Gamma(\alpha)} \int_0^t (t-s)^{\alpha-1} \lambda \big((1+\beta) - Y_s\big) ds + \frac{1}{\Gamma(\alpha)} \int_0^t (t-s)^{\alpha-1} \lambda \sqrt{\frac{1+\beta^2}{\lambda^* \mu(1+\beta)}} \sqrt{Y_s} dB_s, $$
where $(W,B)$ is a correlated bi-dimensional Brownian motion with 
$$ d\langle W,B \rangle_t = \frac{1-\beta}{\sqrt{2(1+\beta^2)}} dt . $$
\noindent Furthermore, the process $Y_t$ has H\"older regularity $\alpha-1/2-\varepsilon$ for any $\varepsilon>0$.
\end{theorem}

\begin{remark}\label{remarkConv}
Theorem \ref{finalHeston} states the convergence in the sense of finite dimensional laws and not in Skorokhod topology. The latter does not hold in general. Nevertheless, we have the convergence for the Skorokhod topology of the integrated price 
$$ \int_0^t \sqrt{\frac{1-a_T}{\mu T ^\alpha}} P_{sT}^T ds $$
to $\int_0^t P_s ds$. Such convergence also holds for the rescaled microscopic price itself under the additional assumption $\varphi_1 = \varphi_2$.
\end{remark}

\noindent Compared to Theorem \ref{Heston1}, the only significant difference in the limiting dynamic here is the kernel $(t-s)^{\alpha-1}$ appearing in the two integrals in the volatility process $Y_t$. Such kernel is similar to that which allows to define a fractional Brownian motion. Indeed, recall  
that a fractional Brownian motion $W^H$ with Hurst parameter $H\in(0,1)$ can be built through the Mandelbrot-van Ness representation:
\begin{equation}
\label{vanness}
W_t^H = \frac{1}{\Gamma(H+1/2)} \int_{-\infty}^0 \big((t-s)^{H-\frac{1}{2}} - (-s)^{H-\frac{1}{2}}\big)dW_s  + \frac{1}{\Gamma(H+1/2)} \int_0^t (t-s)^{H-\frac{1}{2}} dW_s.
\end{equation}
Thus, the tail exponent $\alpha$ in Theorem \ref{finalHeston} corresponds to a Hurst parameter $\alpha-1/2$. Our $\alpha$ belonging to $(1/2,1)$ and in practice being close to $1/2$, the Hurst parameter associated to our limiting volatility is (much) smaller than $1/2$. Therefore, the volatility trajectories are much rougher than that of a Brownian motion and this is why we call our process rough Heston model.\\

\noindent Hence, we have finally shown that when put together in a simple but sufficiently realistic framework, Properties $i$, $ii$, $iii$ and $iv$, which are obvious stylized facts of market microstructure, lead to rough volatility and leverage effect. To our knowledge, this is the first result explaining from an agent-based point of view (although in reduced form) the rough stochastic nature of volatility and in addition leverage effect.\\

\noindent The proof of Theorem \ref{finalHeston} is given in Section \ref{proof4}. As for Theorem \ref{Heston1} it is based on a result on general multidimensional Hawkes processes (but here with heavy tail) which we explain in the next section.

\subsection{Convergence of heavy-tailed nearly unstable multidimensional Hawkes processes} 

We give in this section a general result for the asymptotic behavior of heavy-tailed nearly unstable multidimensional Hawkes processes. This result will be the key to the proof of Theorem \ref{finalHeston}. We consider the same setting as in Section \ref{setting} but we work here under Assumption \ref{fatTail}. This will imply that the result we can obtain here is slightly weaker than that of Theorem \ref{Convergence1}. In particular the sequence of intensities is typically not tight and thus cannot converge. However, the same kind of non-rigorous computations as in Section \ref{intuition1} still enables us to obtain intuition about the result as explained below.

\subsection{Intuition for the result and theorem}

As in Section \ref{intuition1}, we consider a suitable renormalization of the intensity, namely we work with the process
$$C_t^T = \frac{1-a_T}{\mu_T} \lambda_{tT}^T,~~t \in [0,1].$$
Remark that in the setting of Section \ref{intuition1}, the intensity is multiplied by $1/T$. This can be done since under Assumption \ref{smallTail} the factor $(1-a_T)/\mu_T$ is of order $1/T$. This is no longer the case under Assumption \ref{fatTail}.\\

\noindent Following the same computations as in Section \ref{intuition1}, we obtain
$$v_i^*.C_t^T = (1-a_T) (v_i^*.\mathbf{1}) + (v_i^*.\mathbf{1}) \int_0^t \rho_i^T(t-s)ds + \int_0^{t} \rho_i^T(t-s) (v_i^*.d\tilde{M}_s^T),$$  
where $\rho_i^T= T(1-a_T) \psi_i^T(T.)$ and $\tilde{M}_t^T=M_{tT}^T/(T\mu_T)$, which is a martingale such that 
$\mathbb{E}[\langle\tilde{M}^T ,\tilde{M}^T\rangle_t]$ is bounded. Hence we need to study the behavior of $\rho_i^T$.\\

\noindent In the same way as in Section \ref{intuition1}, using its Laplace transform we get that $\rho_i^T$ should vanish as $T$ goes to infinity for $i\geq 2$. For $i=1$, we have
$$\hat{\rho}_1^T(z) = \int_0^\infty \rho_1^T(x) e^{-zx} dx = (1-a_T) \hat{\psi}_1^T(z/T) = (1-a_T) \frac{a_T \hat{\lambda}_1(z/T)}{1-a_T \hat{\lambda}_1(z/T)}.
$$
Then, integrating by parts and using that $\|\lambda_1\|=1$, we get
$$\hat{\lambda}_1(z)= \int_0^\infty \lambda_1(x) e^{-zx} dx = 1 - z \int_0^\infty \int_x^\infty \lambda_1(u)du  e^{-zx} dx.$$
Therefore,
$$\hat{\lambda}_1(z)= 1 - z^\alpha \int_0^\infty (\frac{x}{z})^\alpha \int_{x/z}^\infty \lambda_1(u)du x^{-\alpha} e^{-x} dx.$$
Hence, using Assumption \ref{fatTail} together with the dominated convergence theorem we obtain
$$ \hat{\lambda}_1(z) = 1 - \frac{C}{\alpha} \Gamma(1-\alpha) z^\alpha + \underset{z \rightarrow 0}{o}(z).$$
From this, we easily deduce that for $z>0$,
$$\hat{\rho}_1^T(z){\rightarrow} \frac{\lambda}{\lambda+ z^\alpha},$$
which is the Laplace transform of the Mittag-Leffler density function $f^{\alpha,\lambda}$ defined in Appendix \ref{mittag}. Consequently, using the same arguments as in Section \ref{intuition1}, we get that $C_t^T$ should essentially satisfy
$$ C_t^T \underset{T \rightarrow \infty}{\rightarrow}  \frac{1}{e_1^*.v_1} Y_t e_1,$$
where $Y$ is solution of the following rough stochastic differential equation:
$$ Y_t = (v_1^*.\mathbf{1}) F^{\alpha,\lambda}(t) + \frac{1}{\sqrt{\mu \lambda^*}} \sqrt{\frac{e_1^*.v_1^2}{e_1^*.v_1}} \int_0^t f^{\alpha,\lambda}(t-s)\sqrt{Y_s} dB_s,$$ with $\displaystyle F^{\alpha,\lambda}(t)=\int_0^t f^{\alpha,\lambda}(s)ds$. In fact, this last equation is equivalent to that of a rough Cox-Ingersoll-Ross process:
$$ Y_t = \frac{1}{\Gamma(\alpha)} \int_0^t (t-s)^{\alpha-1} \lambda ( v_1^*.\mathbf{1} - Y_s) ds +  \frac{1}{\Gamma(\alpha)}  \int_0^t (t-s)^{\alpha-1} \frac{\lambda}{\sqrt{\mu \lambda^*}} \sqrt{\frac{e_1^*.v_1^2}{e_1^*.v_1}}  \sqrt{Y_s} dB_s,$$ see Proposition \ref{uniqueness}.\\

\noindent Thus, the preceding computations seem to indicate that in the heavy tail case, the renormalized intensity process should converge to a rough Cox-Ingersoll-Ross process. Contrary to the light tail case, this intuition is actually not correct in general when the kernel matrix has a slowly decreasing behavior. However, it still holds provided we consider the integrated intensity instead of the intensity itself. We now give the rigorous result.\\

\noindent For $t \in [0,1]$, let us define
$$X_t^T= \frac{1-a_T}{T^\alpha \mu} N_{tT}^T,~~\Lambda_t^T= \frac{1-a_T}{T^\alpha \mu} \int_0^{tT} \lambda_{s}^T ds,~~Z_t^T = \sqrt{\frac{T^\alpha \mu}{1-a_T}} (X_t^T - \Lambda_t^T).$$
We have the following theorem.
\begin{theorem}\label{Convergence2}
Under the setting of Section \ref{setting} together with Assumption \ref{fatTail}, the process $(\Lambda_t^T, X_t^T, Z_t^T)_{t \in [0,1]}$ converges in law for the Skorokhod topology to $(\Lambda, X, Z)$ where
$$\Lambda_t = X_t =  \frac{1}{e_1^*.v_1} (\int_0^t Y_s ds) e_1$$
and for $1\leq i\leq d$,
$$Z_t^i = \int_0^t \sqrt{ \frac{e_{1,i} }{e_1^*.v_1} Y_s} dB_s^i,$$
where $(B^1,..,B^d) $ is a $d$-dimensional Brownian motion
and $Y$ is the unique solution of the following rough stochastic differential equation:
$$ Y_t = \frac{1}{\Gamma(\alpha)} \int_0^t (t-s)^{\alpha-1} \lambda ( v_1^*.\mathbf{1} - Y_s) ds +  \frac{1}{\Gamma(\alpha)}  \int_0^t (t-s)^{\alpha-1} \frac{\lambda}{\sqrt{\mu \lambda^*}} \sqrt{\frac{e_1^*.v_1^2}{e_1^*.v_1}}  \sqrt{Y_s} dB_s,$$
with $$ B = \frac{1}{\sqrt{e_1^*.v_1^2}} \sum_{i=1}^{d} \sqrt{e_{1,i} v_{1,i}^2} B^i.$$
Furthermore, $Y$ has H\"older regularity $\alpha - \frac{1}{2} - \varepsilon$ for any $\varepsilon>0$.
\end{theorem}
\noindent The rigorous proof of Theorem \ref{Convergence2} is given in Section \ref{proof3}.

\subsubsection{Application to our microscopic model}
As for Theorem \ref{Convergence1}, Theorem \ref{Convergence2} has an immediate corollary which will be crucial in the proof of Theorem \ref{finalHeston}. 
Let us consider a bi-dimensional Hawkes processes sequence $N^T = (N^{T,+},N^{T,-})$ with intensity $\lambda^T = (\lambda^{T,+},\lambda^{T,-})$ as in Assumption \ref{assumptionStructure}. We have the following result.
\begin{corollary} 
\label{corol2}
Under Assumptions \ref{assumptionStructure} and \ref{fatTail}, the process $(\Lambda_t^T, X_t^T, Z_t^T)_{t \in [0,1]}$  converges in law for the Skorokhod topology to $(X,X,Z)$ where
$$X_t=\frac{1}{\beta+1} \int_0^t Y_s ds \begin{pmatrix} 1 \\ 1\end{pmatrix},~~Z_t = \int_0^t \sqrt{ \frac{1}{\beta + 1} Y_s} \begin{pmatrix}  dB_s^1 \\ dB_s^2\end{pmatrix},$$ 
where $(B^1,B^2) $ is a $bi$-dimensional Brownian motion
and $Y$ is the unique solution of the following rough stochastic differential equation:
$$ Y_t = \frac{1}{\Gamma(\alpha)} \int_0^t (t-s)^{\alpha-1} \lambda ((1+\beta) - Y_s) ds + \frac{1}{\Gamma(\alpha)} \int_0^t (t-s)^{\alpha-1} \lambda \sqrt{\frac{1+\beta^2}{\lambda^* \mu(1+\beta)}} \sqrt{Y_s} dB_s, $$
with $$B = \frac{B^1 + \beta B^2}{\sqrt{1+\beta^2}}.$$
\end{corollary}

\section{Proofs}\label{proofs}
From now on, $c$ denotes a positive constant independent of $T$ that may vary from line to line.
\subsection{Proof of Theorem \ref{Convergence1}}\label{proof1}
In this proof, which is quite inspired by \cite{jaisson2015limit}, the notations defined in Section \ref{intuition1} are in force. We start with a lemma often used in the sequel.
\subsubsection{A useful lemma}
We have the following result.
\begin{lemma} \label{Result}
Let $f^T: \mathbb{R}_+ \rightarrow \mathbb{R} $ be a sequence of measurable functions such that for some $c>0$ and any $x_1\geq 0$, $x_2\in\mathbb{R}$, $x_3\geq 0$, $x_4\geq 0$ and $T>0$:\\
  
\noindent $ a)~f^T \in \mathbb{L}^1(\mathbb{R}_+) \cap \mathbb{L}^2(\mathbb{R}_+)\text{ and }\int_{x \geq 0} |f^T(x)|^2 dx \underset{T \rightarrow \infty}{\rightarrow} 0,$\\
\noindent $ b)~|f^T(x_1)|\leq c,$\\
\noindent $ c)~|\hat{f}^T(x_2)| \leq c (1 \wedge \frac{1}{|x_2|}),$\\
\noindent $ d)~|f^T(x_3)-f^T(x_4)|\leq c T |x_3-x_4|.$\\

\noindent Then, under the setting of Section \ref{setting} together with Assumptions \ref{smallTail} and \ref{regularity}, the process
$$ (\int_0^t f^T(t-s) d\overline{M}_s^T)_{t \in [0,1]}  $$
converges to zero in probability as $T$ goes to infinity, uniformly over compact sets (u.c.p.).
\end{lemma}
\noindent The proof of Lemma \ref{Result} is given in Appendix \ref{abundant}.

\subsubsection{Convergence of $v_i^*.C^T$ for $i \in \{2,...,d\}$}
We now consider the convergence of $C^T$ on the vector space $span(v_2,..,v_d)$. The following proposition holds.
\begin{prop}\label{Convergencevi}
Let $2 \leq i \leq d$. Under the setting of Section \ref{setting} together with Assumptions \ref{smallTail} and \ref{regularity}, ${v_i}^*.C^T$ converges u.c.p. to zero as $T$ goes to infinity.
\end{prop}

\noindent \textsc{Proof}:\\ 

\noindent Recall first Equation \eqref{coordonateRescaledIntensity}:  
$$ v_i^*. C_t^T = \frac{\mu}{T} (v_i^*.\mathbf{1}) + \mu (v_i^*. \mathbf{1}) \int_0^t \psi_i^T(T(t-s)) ds + \int_0^t  \psi_i^T(T(t-s)) (v_i^*.d\overline{M}_s^T).  $$
To get the result, it is therefore enough to show that the family of functions $(\psi_i^T(T.))_{T>0}$ satisfies the four points of Lemma \ref{Result}. Point $b)$ is easily obtained from the fact that $v_i^*.\psi^T = \psi_i^T v_i^*$ together with the uniform boundedness of $\psi^T$ due to Assumption \ref{regularity}.\\

\noindent Now remark that from Assumption \ref{regularity}, we deduce that $\lambda_i(x)$ tends to zero as $x$ goes to infinity. Then, using integration by parts on the Fourier transform of $\lambda_i$ together with Assumption \ref{regularity}, we obtain
\begin{equation} \label{L2}
|\hat{\lambda}_i(\omega)| \leq \big( (|\lambda_i(0)| + \int_0^\infty | \lambda_i'(x) | dx ) \frac{1}{ |\omega|}\big)  \wedge  \| \lambda_i \|_1. 
\end{equation}
Point $c)$ follows using that 
$$|\hat{\psi}_i^T(T.)(\omega)| = \frac{|a_T  \hat{\lambda}_i(\omega/T)|}{|T\big(1-a_T  \hat{\lambda}_i(\omega /T)\big)|} \leq \frac{|a_T  \hat{\lambda}_i(\omega/T)|}{T(1-\| \lambda_i \|_1)} \leq  c(1\wedge \frac{1}{|\omega|}).$$ 

\noindent We also obtain from the previous inequality that $\hat{\psi}_i^T(T.)$ is square-integrable and so is $\psi_i^T(T.)$. Moreover by Parseval equality, we have
$$
\int_{x \geq 0} |\psi_i^T(Tx)|^2 dx = \frac{1}{2 \pi} \int_{\omega \in \mathbb{R}} |\hat{\psi}_i^T(T.)(\omega)| ^2 d\omega
\leq c \int_{\omega \in \mathbb{R}} \frac{ |\hat{\lambda}_i(\omega/T)|^2}{T^2(1- \|\lambda_i\|_1 )^2} d\omega \leq \frac{c}{T} \int_{z \in \mathbb{R}}   |\hat{\lambda}_i(z)|^2 dz.
$$
Since $\hat{\lambda}_i$ is square-integrable, the right hand side of the last inequality tends to zero and thus $a)$ is obtained. \\

\noindent Finally $d)$ is shown using that $\psi_i^T = a_T \lambda_i + a_T \lambda_i*\psi_i^T$ to write
\begin{align*}
|(\psi_i^T)'(Tx)| &= T|a_T  \lambda_i'(Tx) + a_T  (\lambda_i'*\psi_i^T)(Tx) + a_T \lambda_i(0) \psi_i^T(Tx)| \\
&\leq T( \|\lambda_i'\|_{\infty} + \|\lambda_i'\|_{1} \|\psi_i^T\|_{\infty}+|\lambda_i(0)| \|\psi_i^T\|_{\infty}).
\end{align*}
\qed

\subsubsection{{Convergence of $v_1^*.C^T$}}
We have just shown that $v_i^*.C^T$  tends to zero for $i \in \{2,...,d\}$. The fact that $\|\lambda_i\|_1 < 1$ for $i \in \{2,...,d\}$ was crucial in order to obtain this result. We now treat the term $v_1^*.C^T$, recalling that $\|\lambda_1\|_1 = 1$. We have the following result.

\begin{prop} \label{v1Conv}
Under the setting of Section \ref{setting} together with Assumptions \ref{smallTail} and \ref{regularity}, the process $(v_1^*.C_t^T, B_t^{T})_{t \in [0,1]}$ converges in law for the Skorokhod topology to $(X , B)$ where $B$ is a Brownian motion and $X$ satisfies the following Cox-Ingersoll-Ross dynamic:
$$ dX_t = \frac{\lambda}{m}(\frac{\mu}{\lambda}(v_1^*.\mathbf{1}) - X_t) dt + \frac{1}{m} \sqrt{\frac{e_1^*.v_1^2}{e_1^*.v_1}} \sqrt{X_t} dB_t,~~X_0 = 0.$$
\end{prop}

\noindent \textsc{Proof}: 
\paragraph{Rewriting $v_1^*.C^T$} Let 
$$ S_t^T =  \sum_{i=2}^d (e_i^*.C_t^T) (e_i^*.v_1^2) + \big((v')^*.C_t^T\big) (e_1^*.v_1^2).$$
From Proposition \ref{Convergencevi}, we get that $S_t^T$ tends u.c.p. to zero. We have
$$ (v_1^2)^*.C_t^T = S_t^T + \frac{e_1^*.v_1^2}{e_1^*.v_1} v_1^*.C_t^T,$$ which together with \eqref{v1RescaledIntensity} leads to the following expression for $v_1^*.C^T$:
$$ v_1^*.C_t^T =  \frac{\mu}{T} (v_1^*.\mathbf{1}) +   \mu (v_1^*.\mathbf{1}) \int_0^t \psi_1^T(Ts) ds + \int_0^{t} \psi_1^T\big(T(t-s)\big) \sqrt{S_s^T + \frac{e_1^*.v_1^2}{e_1^*.v_1} (v_1^*.C_s^T) } dB_s^{T}.$$

\paragraph{Convergence of $\psi_1^T(T.)$ }For $x\geq 0$, let us define 
$$f^T(x)= \psi_1^T(Tx) - \frac{1}{m} \exp(-\frac{\lambda x}{m}).$$ We have seen in Section \ref{intuition1} that $f^T(x)$ should be close to zero as $T$ goes to infinity. More precisely, we have the following proposition whose proof is given in \cite{jaisson2015limit}:
\begin{prop} \label{propf2}
Under the setting of Section \ref{setting} together with Assumptions \ref{smallTail} and \ref{regularity}, $f^T$ satisfies Properties $a)$, $b)$, $c)$ and $d)$ of Lemma \ref{Result}.
\end{prop}

\paragraph{The Cox-Ingersoll-Ross like dynamic of $v_1^*.C^T$} We can write
$$ v_1^*.C_t^T =  R_t^T +   \frac{\mu}{m} (v_1^*.\mathbf{1}) \int_0^t  \exp(-\frac{\lambda s}{m}) ds  + \frac{1}{m} \sqrt{\frac{e_1^*.v_1^2}{e_1^*.v_1}} \int_0^{t}   \exp(-\frac{\lambda (t-s)}{m}) \sqrt{ v_1^*.C_s^T } dB_s^{T},$$
with
\begin{align}
R_t^T &=   \frac{\mu}{T} (v_1^*.\mathbf{1}) +   \mu  (v_1^*.\mathbf{1}) \int_0^t f^T(s) ds + \int_0^{t} f^T(t-s)( v_1^*.d\overline{M}^T_s)\label{defRT} \\
&+ \frac{1}{m} \int_0^{t}   \exp(-\frac{\lambda (t-s)}{m}) \big(\sqrt{ S_s^T + \frac{e_1^*.v_1^2}{e_1^*.v_1} (v_1^*.C_s^T) } - \sqrt{ \frac{e_1^*.v_1^2}{e_1^*.v_1} (v_1^*.C_s^T) }\big) dB_s^{T}.\nonumber
\end{align}
Then, using integration by parts, we get that 
$$\int_0^{t} \exp(-\frac{\lambda (t-s)}{m}) \sqrt{ v_1^*.C_s^T } dB_s^{T}$$
is equal to
$$\int_0^{t} \sqrt{ v_1^*.C_s^T } dB_s^{T}  - \frac{\lambda}{m} \int_0^{t} \int_0^{s} \exp(-\frac{\lambda (s-u)}{m}) \sqrt{ v_1^*.C_u^T } dB_u^{T} ds.$$
This can be rewritten
$$\int_0^{t} \sqrt{ v_1^*.C_s^T } dB_s^{T}  - \lambda\sqrt{\frac{e_1^*.v_1}{e_1^*.v_1^2}} \int_0^{t}  v_1^*.C_s^T - R_s^T - \frac{\mu}{\lambda} (v_1^*.\mathbf{1}) \big(1 - \text{exp}(-\frac{\lambda s}{m})\big)ds.$$
Consequently,
\begin{equation}\label{convv1}
 v_1^*.C_t^T =  U_t^T + \int_0^t \frac{\lambda}{m} \big( \frac{\mu}{\lambda} (v_1^*.\mathbf{1})- v_1^*.C_s^T\big)ds + \frac{1}{m} \sqrt{\frac{e_1^*.v_1^2}{e_1^*.v_1}} \int_0^t \sqrt{v_1^*.C_s^T} dB_s^{T},\end{equation}
with
$$U_t^T= R_t^T + \frac{\lambda}{m} \int_0^t R_s^T ds.$$
\paragraph{Convergence of $U^T$} We now show that $U^T$ converges u.c.p. to zero. This vanishing behavior comes from that of $f^T$ and $S^T$. Of course it is enough to prove that $R^T$ converges to zero. From Proposition \ref{propf2} together with Lemma \ref{Result}, it is obvious that the first three terms in \eqref{defRT} tend to zero. We now treat the last term.\\

\noindent First, remark that
 $$|\sqrt{ S_s^T + \beta v_1.C_s^T } - \sqrt{ \beta v_1.C_s^T }| \leq\sqrt{ |S_s^T|},$$ which tends to zero as $T$ goes to infinity thanks to Proposition \ref{Convergencevi}. Furthermore, observe that since $\langle M^T,M^T \rangle = diag(\int_0^. \lambda^T)$ and $\lambda^T \geq \mu \mathbf{1}$, we have
\begin{equation} \label{crochet}
{\mathbb{E}} \Big{[} \int_0^{tT} \frac{(v_1^2)^*.dM_s^T}{T (v_1^2)^*.\lambda_s^T} \Big{]}^2 = {\mathbb{E}} \Big{[} \int_0^{tT} \frac{(v_1^4)^*.\lambda_s^T ds}{T^2 \big((v_1^2)^*.\lambda_s^T\big)^2 } \Big{]} \leq   {\mathbb{E}} \Big{[} \int_0^{tT} \frac{(v_1^4)^*.\lambda_s^T ds}{T^2 \mu \big((v_1^4)^*.\lambda_s^T\big) } \Big{]} .
\end{equation}
We get that this is smaller than $c/T$ and consequently goes to zero. Therefore, $B^{T}$ is a sequence of martingales with 
bounded jumps whose quadratic variation, given by
$$[B^{T},B^{T}]_t = t + \int_0^{tT} \frac{(v_1^2)^*.dM_s^T}{T (v_1^2)^*.\lambda_s^T},$$ tends to identity. Using Theorem VII-3.11 in \cite{jacod2013limit}, this implies that $B^T$ converges in law towards a Brownian motion $B$ for the Skorokhod topology. From Theorem 2.6 in \cite{jakubowski1989convergence}, the convergence to zero of 
$$ \frac{1}{m} \int_0^{t}   \exp(-\frac{\lambda (t-s)}{m}) \big(\sqrt{ S_s^T + \frac{e_1^*.v_1}{e_1^*.v_1^2} (v_1^*.C_s^T) } - \sqrt{ \frac{e_1^*.v_1}{e_1^*.v_1^2} (v_1^*.C_s^T) }\big) dB_s^{T}$$
follows and finally we get that $U^T$ tends to zero.

\paragraph{End of the proof of Proposition \ref{v1Conv}}
We have that $v_1^*.C_t^T$ can be written as in \eqref{convv1} and furthermore
$(B^{T},U^T)$ converges in law for the Skorokhod topology to $(B,0)$. Proposition \ref{v1Conv} readily follows from Theorem 5.4 in \cite{kurtz1991weak}.
\qed

\subsubsection{End of proof of Theorem \ref{Convergence1}}
Decomposing $C^T$ in the basis $(e_1,...,e_d)$:
$$ C_t^T = \sum_{i=2}^d (e_i^*.C_t^T) e_i + \big((v')^*.C_t^T\big) e_1 +\frac{1}{e_1^*.v_1} (v_1^*.C_t^T) e_1,$$
we immediately obtain Theorem \ref{Convergence1} from Proposition \ref{Convergencevi} together with Proposition \ref{v1Conv}.

\subsection{Proof of Theorem \ref{Heston1}}\label{proof2}
\subsubsection{Convenient rewriting of $P^T$}
We start by writing conveniently our rescaled price $P_{tT}^T/T$. We have
$$
\frac{1}{T}P_{tT}^T= \frac{N_{tT}^{T,+}-N_{tT}^{T,-}}{T}= \int_0^{tT} \frac{dM_s^{T,+}- dM_s^{T,-}}{\sqrt{T(\lambda_s^{T,+} + \lambda_s^{T,-})}} \sqrt{\frac{\lambda_s^{T,+} + \lambda_s^{T,-}}{T}} + \int_0^{tT}  \frac{\lambda_s^{T,+} - \lambda_s^{T,-}}{T} ds.
$$
Furthermore,
\begin{align*}
\lambda_t^{T,+} - \lambda_t^{T,-}&= \int_0^t a_T(\varphi_1(t-s) - \varphi_2(t-s)) (dN_s^{T,+} - dN_s^{T,-})\\
&= \int_0^t a_T \lambda_2(t-s) (dM_s^{T,+} - dM_s^{T,-}) + \int_0^t a_T \lambda_2(t-s) (\lambda_s^{T,+} - \lambda_s^{T,-}) ds.
\end{align*}
Thus, from Lemma \ref{hopf}, we obtain 
$$ \lambda_t^{T,+} - \lambda_t^{T,-} = \int_0^t \psi_2^T(t-s) (dM_s^{T,+} - dM_s^{T,-}).$$
Then, using Fubini theorem, we get
$$ \int_0^x  \lambda_s^{T,+} - \lambda_s^{T,-} ds = \int_0^x \big{(} \int_0^{x-s} \psi_2^T(u)du \big{)} (dM_s^{T,+}-dM_s^{T,-}).$$
Hence our rescaled price process $P_{tT}^T/T$ can finally be written
$$\int_0^{t} \sqrt{C_s^{T,+} + C_s^{T,-}} dW_s^{T} - \int_0^{t}  \int_{T(t-s)}^{\infty}  \psi_2^T(u) du (d\overline{M}_s^{T,+}-d\overline{M}_s^{T,-}) + \int_{0}^{\infty}  \psi_2^T(u) du  (\overline{M}_t^{T,+}-\overline{M}_t^{T,-}),$$
with
$$ W_t^T = \int_0^{tT} \frac{dM_s^{T,+}-dM_s^{T,-}}{\sqrt{T (\lambda_s^{T,+}+\lambda_s^{T,-})}},$$
and therefore
\begin{equation}\label{convP}\frac{1}{T}P_{tT}^T = \frac{1}{1-a_T(\|\varphi_1\|_1 -\|\varphi_2\|_2)} \int_0^{t} \sqrt{C_s^{T,+} + C_s^{T,-}} dW_s^{T} - R_t^T,\end{equation}
with 
$$R_t^T=\int_0^{t}  \int_{T(t-s)}^{\infty}  \psi_2^T(u) du (d\overline{M}_s^{T,+}-d\overline{M}_s^{T,-}).$$

\subsubsection{Convergence of $R^T$}
We have the following proposition.
\begin{prop}\label{convR} Under Assumptions \ref{assumptionStructure}, \ref{smallTail} and \ref{regularity}, $R^T$ tends u.c.p. to zero.
\end{prop}

\noindent \textit{Proof.}\\
 
\noindent From Lemma \ref{Result}, it is enough to show that the sequence of functions $$g^T(x)=  \int_{Tx}^{\infty}  \psi_2^T(u) du$$ satisfies Properties $a)$, $b)$, $c)$ and $d)$ of Lemma \ref{Result}. The fact that $b)$ holds is obvious since
 $$ |g^T(z)|  \leq \int_{0}^{+\infty} |\psi_2^T(x)| dx \leq \frac{||\lambda_2||_1}{1-||\lambda_2||_1}.$$ Then we remark that 
 $$\hat{g}^T(z) = \int_{a\geq 0} \psi_2^T(a) \frac{e^{iz a /T } -1} {iz} da,$$ which shows that Property $c)$ holds. Property $d)$ is obtained from the fact that $$|(g^T)'(x)| = T |\psi_2^T(Tx)| \leq c T.$$ Finally, we use Fubini theorem to write 
$$\int_{x \geq 0} |g^T(x)|^2 dx = \int_{x \geq 0; a,b >Tx}  \psi_2^T (a) \psi_2^T (b) da db dx =  \frac{1}{T} \int_{a,b \geq 0} (a \wedge b) \psi_2^T (a) \psi_2^T (b)da db.$$
Consequently,
$$
\int_{x \geq 0} |g^T(x)|^2 dx \leq  \frac{1}{T} \int_{a \geq 0} a|\psi_2^T (a)|  da  \int_{b \geq 0} |\psi_2^T (b)| db \leq \frac{c}{T} \sum_{k \geq 1}   \int_{a\geq 0} a |\lambda_2|^{*k}(a) da.
$$
By recursion, we get that for $k\geq 1$, $$\int_{a\geq 0} a |\lambda_2|^{*k}(a) da = k ||\lambda_2||_1^{k-1} \int_{a\geq 0} a |\lambda_2|(a) da < \infty.$$ 
Eventually $$\int_{x \geq 0} |g^T(x)|^2 dx \leq c/T$$ and $a)$ easily follows.
\qed

\subsubsection{Convergence of $(W^{T},B^{T})$}
In the same way as for the quadratic variation of $B^T$ in the proof of Theorem \ref{Convergence1}, we easily get the following convergence in probability:
$$ [W^{T},W^{T}]_t \underset{T \rightarrow \infty}{\rightarrow} t, \quad  [B^{T},B^{T}]_t \underset{T \rightarrow \infty}{\rightarrow} t.$$
Moreover, we have the following proposition.
\begin{prop}\label{correlfound} Under Assumptions \ref{assumptionStructure}, \ref{smallTail} and \ref{regularity},
$$ [W^{T},B^{T}]_t \underset{T \rightarrow \infty}{\rightarrow} \frac{1-\beta}{\sqrt{2(1+\beta^2)}}t$$
in probability.
\end{prop}
 
 \noindent \textit{Proof.}\\
 
 \noindent Using $[M^T,M^T] = diag(N^T)$, we write
 \begin{align*}
 [W^{T},B^{T}]_t =& \int_{0}^{tT} \frac{dN_s^{T,+}- \beta dN_s^{T,-}}{T \sqrt{\lambda_s^{T,+}+\lambda_s^{T,-}} \sqrt{\lambda_s^{T,+}+\beta^2 \lambda_s^{T,-}}}\\ =&\int_{0}^{t} \frac{C_s^{T,+}- \beta C_s^{T,-}}{ \sqrt{C_s^{T,+}+C_s^{T,-}} \sqrt{C_s^{T,+}+\beta^2 C_s^{T,-}}} ds+\varepsilon_t^T,
 \end{align*}
 with
 $$\varepsilon_t^T= \int_{0}^{tT} \frac{dM_s^{T,+}- \beta dM_s^{T,-}}{T \sqrt{\lambda_s^{T,+}+\lambda_s^{T,-}} \sqrt{\lambda_s^{T,+}+\beta^2 \lambda_s^{T,-}}}.$$
Since $\langle M^T,M^T \rangle = diag(\int_0^. \lambda^T)$ and $\lambda^T \geq \mu \mathbf{1}$, we easily get
  $$
 {\mathbb{E}}[(\varepsilon_t^T)^2] = {\mathbb{E}} \big{[} \int_{0}^{tT} \frac{1}{T^2 (\lambda_s^{T,+}+\lambda_s^{T,-})} \big{]} \leq \frac{1}{2 \mu T} \underset{T \rightarrow \infty}{\rightarrow} 0.
  $$
Furthermore, from Corollary \ref{corol1}, $(C^{T,+}, C^{T,-})$ converges in law for the Skorokhod topology to $\big{(} \frac{1}{\beta+1} X,\frac{1}{\beta+1} X\big{)}$. The set of zeros of a Cox-Ingersoll-Ross process on a finite interval being of Lebesgue measure zero, we deduce that
$$\frac{C_t^{T,+}- \beta C_t^{T,-}}{\sqrt{C_t^{T,+}+C_t^{T,-}} \sqrt{C_t^{T,+}+\beta^2 C_t^{T,-}}}$$
tends u.c.p. to 
$$\frac{1-\beta}{\sqrt{2(1+\beta^2)}}.$$ Thus we deduce that
$$ [B^{T},W^{T}]_t \underset{T \rightarrow \infty}{\rightarrow} \frac{1-\beta}{\sqrt{2(1+\beta^2)}}t.$$
\qed 

\subsubsection{End of the proof of Theorem \ref{Heston1}}

\noindent Consider \eqref{convP}. From Proposition \ref{convR}, $R^T$ tends to zero. Then using 
Theorem VII-3.11 in \cite{jacod2013limit} together with Proposition \ref{correlfound}, we obtain that $(W^T,B^T)$ converges in law for the Skorokhod topology to a correlated bi-dimensional Brownian motion $(W,B)$ such that
$$\langle W,B \rangle_t = \frac{1-\beta}{\sqrt{2(1+\beta^2)}} t.$$ Furthermore, from Corollary \ref{corol1} we get that
$(\sqrt{C^{T,+} + C^{T,-}},B^T) $ converges in law for the Skorokhod topology to $(\sqrt{\frac{2}{\beta+1} X},B)$, where $X$ is a Cox-Ingersoll-Ross process driven by $B$ and defined in Corollary \ref{corol1}. Using Theorem 2.6 in \cite{jakubowski1989convergence}, we deduce that $$\int_0^{t} \sqrt{C_s^{T,+} + C_s^{T,-}} dW_s^{T}$$ 
converges in law for the Skorokhod topology to $$\int_0^{t} \sqrt{\frac{2 X_s}{1+\beta}} dW_s,$$ which ends the proof.

\subsection{Proof of Theorem \ref{Convergence2}}\label{proof3}
We now give the proof of our theorem on the convergence of general nearly unstable Hawkes processes with heavy tail. This proof is quite inspired from  \cite{jaisson2016rough}.
\subsubsection{{C-tightness of $(\Lambda^T,X^T,Z^T)$ }}
We have the following proposition.
\begin{prop}\label{cTight}
Under the setting of Section \ref{setting} together with Assumption \ref{fatTail}, the sequence $(\Lambda^T,X^T,Z^T)$ is C-tight
and
$$ \underset{t \in [0,1]}{\sup} \|\Lambda_t^T-X_t^T\| \underset{T \rightarrow \infty}{\rightarrow} 0$$
in probability. Moreover if $(X,Z)$ is a possible limit point of $(X^T,Z^T)$, then $Z$ is a continuous martingale with $[Z,Z] = diag(X)$.
\end{prop}

\noindent \textsc{Proof}:
\paragraph{C-tightness of $X^T$ and $\Lambda^T$} Recall that as in \eqref{intensity3}, we can write
$$\lambda_t^T = \mu_T \mathbf{1} +\mu_T \int_0^t \psi^T(t-s)ds.\mathbf{1}+ \int_0^t \psi^T(t-s).dM_s^T.$$
Using that $\displaystyle\int_0^. (f*g) = (\int_0^.f)*g$, we get 
$$ {\mathbb{E}}[N_T^T] = {\mathbb{E}}[\int_0^T \lambda_s^T ds] = T \mu_T \mathbf{1} + \mu_T \int_0^T s \psi^T(T-s) ds.\mathbf{1}.$$
Consequently,
$$
\mathbf{1}^*.\mathbb{E}[N_T^T] = T \mu_T d + \mu_T \mathbf{1}^*.(\int_0^T s \psi^T(T-s) ds).\mathbf{1}$$ and therefore
$$\mathbf{1}^*.\mathbb{E}[N_T^T] \leq  c T \mu_T \big(1 + {\cal S}(\int_0^\infty \psi^T(s) ds)\big) \leq  c \frac{T \mu_T}{ 1-a_T} \leq c T^{2\alpha}.$$
Thus, we obtain that $$\mathbb{E}[X_1^T]  = \mathbb{E}(\Lambda_1^T)\leq c.$$ Each component of $X^T$ and $\Lambda^T$ being increasing, we deduce the tightness of each component of $(X^T, \Lambda^T)$. Furthermore, the maximum jump size of $X^T$ and $\Lambda^T$ being $\frac{1-a_T}{T^\alpha \mu}$ which goes to zero, the C-tightness of $(X^T,\Lambda^T)$ is obtained from Prop.VI-3.26 in \cite{jacod2013limit}. 

\paragraph{C-tightness of $Z^T$} It is easy to check that
$$ \langle Z^T,Z^T \rangle = diag(\Lambda^T),$$
which is C-tight. From Theorem VI-4.13 in \cite{jacod2013limit}, this gives the tightness of $Z^T$. The maximum jump size of $Z^T$ vanishing as $T$ goes to infinity, we obtain that $Z^T$ is C-tight. 

\paragraph{Convergence of $X^T-\Lambda^T$} We have
$$X_t^T-\Lambda_t^T = \frac{1-a_T}{T^\alpha \mu} M_{tT}^T.$$ 
From Doob's inequality, we get for each component
\begin{align*}
{\mathbb{E}}[\underset{t \in [0,1]}{\sup} |\Lambda_t^T-X_t^T|^2] &\leq c T^{-4\alpha} \mathbb{E}[M_T^T]^2.
\end{align*}
Since $[M^T,M^T] = diag(N^T)$, we deduce
$$
{\mathbb{E}}[\underset{t \in [0,1]}{\sup} |\Lambda_t^T-X_t^T|^2] \leq c T^{-2\alpha}.
$$
This gives the uniform convergence to zero in probability of $X^T-\Lambda^T$.

\paragraph{Limit of $Z^T$}
\noindent Let $(X,Z)$ be a possible limit point of $(X^T,Z^T)$. We know that $(X,Z)$ is continuous and from Corollary IX-1.19 of \cite{jacod2013limit}, $Z$ is a local martingale. Moreover, since
$$ [Z^T,Z^T] = diag(X^T),  $$
using Theorem VI-6.26 in \cite{jacod2013limit}, we get that $[Z,Z]$ is the limit of $[Z^T,Z^T]$ and $[Z,Z] = diag(X)$. By Fatou's lemma, the expectation of $[Z,Z]$ is finite and therefore $Z$ is a martingale.
\qed

\subsubsection{Convergence of $v_i^*.X^T$ for $i \geq 2$}
Here also, we observe a vanishing behavior in the direction of the eigenvectors $v_i$ for $i\geq 2$. More precisely, we have the following result.
\begin{prop} \label{ziro}
Under the setting of Section \ref{setting} together with Assumption \ref{fatTail}, if $X$ is a possible limit point of $X^T$, then for $i\geq 2$ we have
${v_i}^*.X = 0$.
\end{prop}

\noindent \textsc{Proof}: 

\noindent From \eqref{intensity3}, using Fubini theorem together with the fact that $\displaystyle\int_0^. (f*g) = (\int_0^. f)*g$, we get
$$ \int_0^t \lambda_s^T ds = t \mu_T \mathbf{1} + \mu_T \int_0^t s \psi^T(t-s)  ds. \mathbf{1}  + \int_0^t \psi^T(t-s). M_s^T ds.$$
Therefore, for $t\in [0,1]$, we have the decomposition 
\begin{equation} \label{decomp}
\Lambda_t^T =T_1 + T_2 + T_3, 
\end{equation}
with
$$ T_1 = (1-a_T) t u_T \mathbf{1},$$
$$ T_2 = T (1-a_T) u_T \int_0^t s \psi^T\big(T(t-s)\big)  ds. \mathbf{1},$$
$$ T_3 = T^{1-\alpha/2} \sqrt{\frac{1-a_T}{\mu}} \int_0^t  \psi^T\big(T(t-s)\big) . Z_s^T ds ,$$
with $u_T =\mu_T/(\mu T^{\alpha-1})$ tending to one.\\

\noindent Now recall that for $1 \leq i \leq d$,
$$ \psi_i^T = \sum_{k\geq 1} a_T^k \lambda_i^{*k}, \quad \rho_i^T = T(1-a_T)  \psi_i^T(T.),$$
and define 
$$ F_i^T= \int_0^{.} \rho_i^T(s) ds.$$
For $i\geq 2$, using that
$$|F_i^T(t)|\leq \int_0^t T(1-a_T) |\psi_i^T(Ts)| ds \leq (1-a_T) \int_0^\infty |\psi_i^T(s)| ds \leq (1-a_T) \frac{\|\lambda_i\|_1}{1-\|\lambda_i\|_1}.$$
we get the uniform convergence to zero of $F_i^T$. Thanks to this together with integration by parts, we deduce the convergence to zero of $v_i^*.T_2$ 
since
$$v_i^*.T_2 = u_T (v_i^*.\mathbf{1}) \int_0^t F_i^T(s) ds.$$ For $v_i^*.T_3$ we write 
$$v_i^*.T_3 = \frac{1}{\sqrt{ \mu(1-a_T) T^{\alpha}}} \int_0^t F_i^T(t-s) (v_i^*.dZ_s^T).$$
The quadratic variation of $Z^T$ being $\Lambda^T$ which is uniformly bounded in expectation, we have
$$\mathbb{E}[(v_i^*.T_3)^2]\leq \frac{c}{{ \mu (1-a_T) T^{\alpha}}} \int_0^t \big(F_i^T(s)\big)^2 ds.$$
The convergence of $v_i^*.T_3$ to zero follows. Finally, from Proposition \ref{cTight} we have that if $X$ is a limit point of $X^T$, then $X$ is also a limit point of $\Lambda^T$. Therefore, we obtain $v_i^*.X = 0$.\\
\qed

\subsubsection{Convergence of $v_1^*.X^T$}
The term $v_1^*.X^T$ is the non-vanishing one. Indeed, for $(Z,X)$ a possible limit point of $(Z^T,X^T)$, using the same approach as in \cite{jaisson2016rough}, we obtain 
$$ T_2^*.v_1    \underset{T \rightarrow \infty}{\rightarrow} (v_1^*.\mathbf{1}) \int_0^t s f^{\alpha,\lambda}(t-s) ds$$
and
$$ T_3^*.v_1 \underset{T \rightarrow \infty}{\rightarrow}  \frac{1}{\sqrt{\lambda^* \mu}} \int_0^t f^{\alpha,\lambda}(t-s) (v_1^*.Z_s) ds.$$
Then, letting $T$ go to infinity in the decomposition \eqref{decomp} we easily deduce the following proposition.

\begin{prop} 
\label{v1rough}
Under the setting of Section \ref{setting} together with Assumption \ref{fatTail}, if $(Z,X)$ is a possible limit point of $(Z^T,X^T)$, then the process $v_1^*.X$ satisfies the following equation:
$$ v_1^*.X_t =  (v_1^*.\mathbf{1}) \int_0^t s f^{\alpha,\lambda}(t-s) ds + \frac{1}{\sqrt{\lambda^* \mu}} \int_0^t f^{\alpha,\lambda}(t-s) (v_1^*.Z_s) ds.$$  
\end{prop} 

\subsubsection{A first version of Theorem \ref{Convergence2}} We now prove of version of Theorem \ref{Convergence2} where $Y$ is specified in a different way. Let $(X,Z)$ be a possible limit point of $(X^T,Z^T)$. From Proposition \ref{v1rough}, in the same way as the proof of Theorem 3.2 in \cite{jaisson2016rough}, we can show that$$  v_1^*.X_t = \int_0^t Y_s ds, $$
where $Y$ satisfies 
$$ Y_t =  (v_1^*.\mathbf{1}) F^{\alpha,\lambda}(t) + \frac{1}{\sqrt{\lambda^* \mu}} \int_0^t f^{\alpha,\lambda}(t-s) (v_1^*.dZ_s).$$  
Using Proposition \ref{ziro} together with the decomposition of $X_t$ in the orthonormal basis $(e_1,...,e_d)$:
$$ X_t = \sum_{i=2}^d (e_i^*.X_t) e_i + \big((v')^*.X_t\big) e_1 +\frac{1}{e_1^*.v_1} (v_1^*.X_t) e_1,$$
we get
$$X_t = \frac{1}{e_1^*.v_1} (v_1^*.X_t) e_1 = \frac{1}{e_1^*.v_1} (\int_0^t Y_s ds) e_1.$$
From Proposition \ref{cTight}, we have that $$[Z,Z] = diag(X) = \frac{1}{e_1^*.v_1} (\int_0^t Y_s ds)  diag(e_1).$$ Thus we can use Theorem V-3.9 in \cite{revuz1999continuous} to show the existence of a $d$-dimensional Brownian motion $(B^1,...,B^d)$ such that for $1 \leq i \leq d$,
$$ Z_t^i =  \frac{1}{\sqrt{e_1^*.v_1}}  \sqrt{e_{1,i}} \int_0^t \sqrt{Y_s} dB_s^i.$$
Finally, in the same way as the proof of Theorem 3.2 in \cite{jaisson2016rough}, we obtain that $Y$ satisfies 
\begin{equation}\label{defYtemp}
 Y_t =  (v_1^*.\mathbf{1}) F^{\alpha,\lambda}(t) + \sqrt{\frac{e_1^*.v_1^2}{\lambda^* \mu (e_1^*.v_1)}} \int_0^t f^{\alpha,\lambda}(t-s) \sqrt{Y_s} dB_s,\end{equation}
where $B$ is a Brownian motion defined by 
$$ B = \frac{1}{\sqrt{e_1^*.v_1^2}} \sum_{1 \leq i \leq d} \sqrt{e_{1,i}} v_{1,i} B^i.$$
and that $Y$ has H\"older regularity $\alpha - 1/2 - \varepsilon$, for any $\varepsilon>0$.

\subsubsection{End of the proof of Theorem \ref{Convergence2}}
We eventually provide here the proposition showing that from \eqref{defYtemp}, $Y$ can be written under the form of the rough stochastic differential equation given in Theorem \ref{Convergence2} and stating the uniqueness of the solution of this equation. Theorem \ref{Convergence2} follows immediately.
\begin{prop}
\label{uniqueness}
Let $\lambda$, $\nu$, $\theta$ be positive constants, $\alpha \in (1/2,1)$ and $B$ a Brownian motion. The process $V$ is solution of the following rough stochastic differential equation:
\begin{equation}
\label{SDEMittag}
V_t =  \theta F^{\alpha,\lambda}(t) + \nu \int_0^t f^{\alpha,\lambda}(t-s) \sqrt{V_s} dB_s
\end{equation}
if and only if it is solution of
\begin{equation}
\label{SDEFrac}
V_t = \frac{1}{\Gamma(\alpha)} \int_0^t (t-s)^{\alpha-1} \lambda (\theta - V_s) ds +  \frac{\lambda \nu}{\Gamma(\alpha)}  \int_0^t (t-s)^{\alpha-1} \sqrt{V_s} dB_s.
\end{equation}
Furthermore, both equations admit a unique strong solution.
\end{prop}
\noindent \textsc{Proof}: \\ 

\noindent The existence of a solution to \eqref{SDEMittag} has already been proved deriving \eqref{defYtemp}. Let $V$ be a solution to \eqref{SDEMittag} and write
$$ K = I^{1-\alpha}V,$$
where $I^{1-\alpha}$ is the fractional integral operator of order $(1-\alpha)$, see Appendix \ref{defInt}. Using stochastic Fubini theorem, see for example \cite{veraar2012stochastic}, and integration by parts, we get
$$ K_t = \theta \int_0^t I^{1-\alpha} f^{\alpha,\lambda}(u) du + \nu \int_0^t I^{1-\alpha}f^{\alpha,\lambda}(t-u) \sqrt{V_u} dB_u.$$
Moreover, since $I^{1-\alpha}f^{\alpha,\lambda}(t) = \lambda\big(1-F^{\alpha,\lambda}(t)\big)$, see Appendix \ref{mittag}, using stochastic Fubini theorem, we obtain
$$K_t= \lambda \theta \int_0^t \big(1-F^{\alpha,\lambda}(u)\big) du + \nu \lambda \int_0^t \sqrt{V_u} dB_u -\lambda  \int_0^t \nu \int_0^s \sqrt{V_u} f^{\alpha,\lambda}(s-u) dB_u ds.$$
Hence, 
$$
K_t= \lambda \theta \int_0^t \big(1-F^{\alpha,\lambda}(u)\big) du + \nu \lambda \int_0^t \sqrt{V_u} dB_u -\lambda \int_0^t \big(V_s - \theta F^{\alpha,\lambda}(s)\big) ds$$ and finally
$$K_t=\lambda \int_0^t (\theta - V_u) du + \lambda \nu \int_0^t \sqrt{V_u} dB_u.$$
Now recall that we have 
$$V_t=  D^{1-\alpha}K_t,$$ where the fractional differentiation operator $D^{1-\alpha}$ is defined in Appendix \ref{defInt}.
Thus we get
$$V_t= \frac{1}{\Gamma(\alpha)}\frac{d}{dt}  \int_0^t \lambda \int_0^s (s-u)^{\alpha-1} (\theta - V_u) d_u ds + \frac{1}{\Gamma(\alpha)}\frac{d}{dt}  \int_0^t \lambda \nu \int_0^s (s-u)^{\alpha-1} \sqrt{V_u} dB_u ds$$ and finally, again from Fubini theorem,
$$V_t=  \frac{1}{\Gamma(\alpha)} \lambda \int_0^t (t-u)^{\alpha-1} (\theta - V_u) du+ \frac{1}{\Gamma(\alpha)} \lambda \nu \int_0^t (t-u)^{\alpha-1} \sqrt{V_u} dB_u. 
$$
Therefore $V$ is solution of \eqref{SDEFrac}. Using a straightforward generalization of the main result in \cite{mytnik2015uniqueness}, we deduce the uniqueness of such a solution.\qed

\subsection{Proof of Theorem \ref{finalHeston}} \label{proof4}
First, remark that in the same way as in Section \ref{proof2}, we can write
$$\sqrt{\frac{1-a_T}{T^\alpha \mu}} P_{tT}^T= \frac{1}{1-a_T(\|\varphi_1\|_1-\|\varphi_2\|_1)}(Z_t^{T,+} - Z_t^{T,-} )- R_t^T,$$
with
$$ R_t^T =\int_0^{t}\big(\int_{T(t-s)}^{\infty}  \psi_2^T(u) du\big) (dZ_s^{T,+} - dZ_s^{T,-}).$$
\noindent Using Corollary \ref{corol2}, we deduce that 
$$ \frac{1}{1-a_T(\|\varphi_1\|_1-\|\varphi_2\|_1)}(Z_t^{T,+} - Z_t^{T,-} ) $$
converges in law for the Skorokhod topology to the rough Heston dynamic $P$ defined in Theorem \ref{finalHeston}.\\

\noindent Note that when $\varphi_1 = \varphi_2$, $R^T = 0$. Thus, in this case, we obtain the convergence in law for the Skorokhod topology of the rescaled microscopic price to $P$. For the general case, we can prove the convergence of $R^T$ to zero in the sense of finite dimensional laws as follows. We have
$$ \E[(R_t^T)^2] \leq c \int_0^t  \big(\int_{Ts}^{\infty}  \psi_2^T(u) du\big)^2 ds.$$
Let $G = \sum_{k \geq 1} |\varphi_1 - \varphi_2|^{*k}$.  Note that $G$ is integrable since $\int_0^\infty |\varphi_1 - \varphi_2| < 1$. Hence 
$$ \E[(R_t^T)^2] \leq c \big(T^{-1/2} ( \int_0^\infty G)^2 +  ( \int_{T^{1/2}}^\infty G)^2 \big),$$
which vanishes as $T$ tends to infinity. The result follows.
\qed

\begin{remark}
Note that
$$ \underset{t \in [0,1]}{\sup} |\int_0^t  \int_{Ts}^{\infty}  \psi_2^T(u) du ds| \leq c\big(T^{-1/2} \int_0^\infty G +  \int_{T^{1/2}}^\infty G\big),$$
which vanishes as $T$ goes to infinity. Then, using Fubini theorem, we get that 
$$ \int_0^t R_s^T ds = \int_0^t  \int_0^{t-s} \big( \int_{Tu}^{\infty}  \psi_2^T\big) du (dZ_s^{T,+} - dZ_s^{T,-})$$
converges u.c.p. to zero. Thus, as stated in Remark \ref{remarkConv}, the integrated rescaled microscopic price converges in law for the Skorokhod topology to $\int_0^t P_s ds$.
\end{remark}
\section*{Acknowledgments}

We thank Neil Shephard for inspiring discussions and Jim Gatheral and Kasper Larsen for very relevant comments.

\appendix

\section{{Appendix}}

\subsection{{Proof of Lemma \ref{Result}.}} \label{abundant}

This result has already been proved in  \cite{jaisson2016rough} in dimension one. We need to generalize it for $d \geq 2$. Inspection of the proof in  \cite{jaisson2016rough} shows that the tightness of $$H^T_t=\int_0^t f^T(t-s) d\overline{M}_s^T$$ holds the same way when the dimension is larger than one. So we just need to check the finite dimensional convergence of $Y^T$ to zero. Using that $\langle M^T,M^T \rangle= \int_0^. \lambda^T$, we get
$$
{\mathbb{E}} [\|H_t^T\|_2^2] = \frac{1}{T^2} {\mathbb{E}} \big[ \int_0^{tT} f^T(t- s/T)^2 \sum_{i=1}^d \lambda_{s,i}^T ds \big] =  \frac{1}{T^2}  \int_0^{tT} f^T(t- s/T)^2 \sum_{i=1}^d {\mathbb{E}} [\lambda_{s,i}^T] ds.
$$
Using \eqref{intensity3} together with the fact that $v_i^*.\psi^T = \psi_i^T v_i^*$, we obtain that for any $i \in \{1,..,d\}$ and $s\geq 0$,
$${\mathbb{E}} [v_i^*.\lambda_{s}^T] = \mu (v_i^*.\mathbf{1})\big(1+  \int_0^s \psi_i^T(u) du\big) .$$
Thus
$$ |{\mathbb{E}} [v_i^*.\lambda_{s}^T]|\leq \mu | v_i^*.\mathbf{1} | \big(1+  \sum_{k \geq 1} \int_0^{\infty} a_T^k |\lambda_i|^{*k}(u) du\big) \leq \mu | v_i^*.\mathbf{1} | \frac{1}{1-a_T \|\lambda_i\|_1} \leq c T.$$
Hence for any $i \in \{1,..,d\}$, $ {\mathbb{E}}[\lambda_{s,i}^T] \leq cT$. 
Therefore
$${\mathbb{E}}[\|H_t^T\|_2^2]\leq  c \int_0^{\infty} f^T(s)^2 ds \underset{T \rightarrow \infty}{\rightarrow} 0
$$
and so $H_t^T$ tends in probability to zero giving the finite dimensional convergence of the process.
\subsection{Wiener-Hopf equations}
The following result is used extensively in this work to solve Wiener-Hopf type equations, see for example \cite{bacry2013some}.
\begin{lemma}
\label{hopf}
Let $g$ be a measurable locally bounded function from $\mathbb{R}$ to $\mathbb{R}^d$ and $\phi : \mathbb{R}_+ \rightarrow \cal{M}^{\textbf{d}}(\mathbb{R}) $ be a matrix-valued function with integrable components such that $\mathcal{S}(\int_0^\infty \phi(s) ds) < 1$. Then there exists a unique
locally bounded function $f$ from $\mathbb{R}$ to $\mathbb{R}^d$ solution of
$$f(t) = g(t) + \int_0^t \phi(t-s). f(s) ds,~~t \geq 0$$
given by
$$ f(t) = g(t) + \int_0^t \psi(t-s). g(s) ds,~~t \geq 0, $$
where $\displaystyle\psi= \sum_{k \geq 1} \phi^{*k}$.
\end{lemma}

\subsection{Fractional integrals and derivatives}\label{defInt}
The fractional integral of order $r \in (0,1]$ of a function $f$ is defined by
\begin{equation*} 
I^r f(t) = \frac{1}{\Gamma(r)} \int_0^t (t-s)^{r-1} f(s) ds,  
\end{equation*}
whenever the integral exists. Its fractional derivative of order $r \in [0,1)$ is given by
\begin{equation*} 
D^r f(t) = \frac{1}{\Gamma(1-r)} \frac{d}{dt} \int_0^t (t-s)^{-r} f(s) ds,  
\end{equation*}
whenever it exists.

\subsection{Mittag-Leffler functions}
\label{mittag}
Let $(\alpha,\beta) \in (\mathbb{R}_+^*)^2$. The Mittag-Leffler function $E_{\alpha,\beta}$ is defined for $z \in \mathbb{C}$ by
$$ E_{\alpha,\beta}(z) = \sum_{n \geq 0} \frac{z^n}{\Gamma(\alpha n + \beta)}.$$ 
For $(\alpha,\lambda) \in  (0,1)\times \mathbb{R}_+$, we also define 
$$ f^{\alpha,\lambda}(t) = \lambda t^{\alpha - 1} E_{\alpha,\alpha}(-\lambda t^\alpha),~~t>0,$$
$$ F^{\alpha,\lambda} = \int_0^t f^{\alpha,\lambda}(s) ds,~~t \geq 0.$$
The function $f^{\alpha,\lambda}$ is a density function on $\mathbb{R}_+$ called Mittag-Leffler density function.\\
For $\alpha \in (1/2,1)$, $f^{\alpha,\lambda}$ is square-integrable and its Laplace transform is given for $z \geq 0$ by 
$$\hat{f}^{\alpha,\lambda}(z)= \int_0^\infty f_{\alpha,\lambda}(s) e^{-zs} ds = \frac{\lambda}{\lambda + z^\alpha}.$$
Finally, we can show that 
$$ I^{1-\alpha}f^{\alpha,\lambda}(t) = \lambda \big(1-F^{\alpha,\lambda}(t)\big).$$
Further properties of $f^{\alpha,\lambda}$ and $F^{\alpha,\lambda}$ can be found in \cite{haubold2011mittag,mainardisome,mathai2008special}.

\bibliographystyle{abbrv}
\bibliography{Bibli_EER_final}

\begin{thebibliography}{10}

\bibitem{abergel2014understanding}
F.~Abergel, C.-A. Lehalle, and M.~Rosenbaum.
\newblock Understanding the stakes of high-frequency trading.
\newblock {\em The Journal of Trading}, 9(4):49--73, 2014.

\bibitem{ait2015modeling}
Y.~A{\"\i}t-Sahalia, J.~Cacho-Diaz, and R.~J. Laeven.
\newblock Modeling financial contagion using mutually exciting jump processes.
\newblock {\em Journal of Financial Economics}, 117(3):585--606, 2015.

\bibitem{ait2013leverage}
Y.~Ait-Sahalia, J.~Fan, and Y.~Li.
\newblock The leverage effect puzzle: Disentangling sources of bias at high
  frequency.
\newblock {\em Journal of Financial Economics}, 109(1):224--249, 2013.

\bibitem{almgren2001optimal}
R.~Almgren and N.~Chriss.
\newblock Optimal execution of portfolio transactions.
\newblock {\em Journal of Risk}, 3:5--40, 2001.

\bibitem{bacry2012non}
E.~Bacry, K.~Dayri, and J.-F. Muzy.
\newblock Non-parametric kernel estimation for symmetric hawkes processes.
  application to high frequency financial data.
\newblock {\em The European Physical Journal B}, 85(5):1--12, 2012.

\bibitem{bacry2013modelling}
E.~Bacry, S.~Delattre, M.~Hoffmann, and J.-F. Muzy.
\newblock Modelling microstructure noise with mutually exciting point
  processes.
\newblock {\em Quantitative Finance}, 13(1):65--77, 2013.

\bibitem{bacry2013some}
E.~Bacry, S.~Delattre, M.~Hoffmann, and J.-F. Muzy.
\newblock Some limit theorems for hawkes processes and application to financial
  statistics.
\newblock {\em Stochastic Processes and their Applications}, 123(7):2475--2499,
  2013.

\bibitem{bacry2014estimation}
E.~Bacry, T.~Jaisson, and J.-F. Muzy.
\newblock Estimation of slowly decreasing {H}awkes kernels: Application to high
  frequency order book modelling.
\newblock {\em Quantitative Finance}, 16(8):1179--1201, 2016.

\bibitem{bacry2015hawkes}
E.~Bacry, I.~Mastromatteo, and J.-F. Muzy.
\newblock Hawkes processes in finance.
\newblock {\em Market Microstructure and Liquidity}, 1(01):1550005, 2015.

\bibitem{bauwens2004dynamic}
L.~Bauwens and N.~Hautsch.
\newblock Dynamic latent factor models for intensity processes.
\newblock {\em CORE Discussion Paper}, 2004.

\bibitem{bayer2016pricing}
C.~Bayer, P.~Friz, and J.~Gatheral.
\newblock Pricing under rough volatility.
\newblock {\em Quantitative Finance}, 16(6):887--904, 2016.

\bibitem{bekaert2000asymmetric}
G.~Bekaert and G.~Wu.
\newblock Asymmetric volatility and risk in equity markets.
\newblock {\em Review of Financial Studies}, 13(1):1--42, 2000.

\bibitem{black1976studies}
F.~Black.
\newblock Studies of stock price volatility changes.
\newblock {\em Proceedings of the 1976 Meetings of the American Statistical
  Association}, pages 171--181, 1976.

\bibitem{bollerslev1992arch}
T.~Bollerslev, R.~Y. Chou, and K.~F. Kroner.
\newblock {ARCH} modeling in finance: A review of the theory and empirical
  evidence.
\newblock {\em Journal of Econometrics}, 52(1-2):5--59, 1992.

\bibitem{bollerslev2006leverage}
T.~Bollerslev, J.~Litvinova, and G.~Tauchen.
\newblock Leverage and volatility feedback effects in high-frequency data.
\newblock {\em Journal of Financial Econometrics}, 4(3):353--384, 2006.

\bibitem{bowsher2007modelling}
C.~G. Bowsher.
\newblock Modelling security market events in continuous time: Intensity based,
  multivariate point process models.
\newblock {\em Journal of Econometrics}, 141(2):876--912, 2007.

\bibitem{brennan2012sell}
M.~J. Brennan, T.~Chordia, A.~Subrahmanyam, and Q.~Tong.
\newblock Sell-order liquidity and the cross-section of expected stock returns.
\newblock {\em Journal of Financial Economics}, 105(3):523--541, 2012.

\bibitem{brunnermeier2009market}
M.~K. Brunnermeier and L.~H. Pedersen.
\newblock Market liquidity and funding liquidity.
\newblock {\em Review of Financial studies}, 22(6):2201--2238, 2009.

\bibitem{campbell1992no}
J.~Y. Campbell and L.~Hentschel.
\newblock No news is good news: An asymmetric model of changing volatility in
  stock returns.
\newblock {\em Journal of Financial Economics}, 31(3):281--318, 1992.

\bibitem{chavez2005point}
V.~Chavez-Demoulin, A.~C. Davison, and A.~J. McNeil.
\newblock A point process approach to value-at-risk estimation.
\newblock {\em Quantitative Finance}, 5(2):227--234, 2005.

\bibitem{christie1982stochastic}
A.~A. Christie.
\newblock The stochastic behavior of common stock variances: Value, leverage
  and interest rate effects.
\newblock {\em Journal of Financial Economics}, 10(4):407--432, 1982.

\bibitem{corradi2000reconsidering}
V.~Corradi.
\newblock Reconsidering the continuous time limit of the {GARCH} (1, 1)
  process.
\newblock {\em Journal of Econometrics}, 96(1):145--153, 2000.

\bibitem{dayri2012large}
K.~Dayri and M.~Rosenbaum.
\newblock {Large tick assets: Implicit spread and optimal tick size}.
\newblock {\em Market Microstructure and Liquidity}, 1:1550003, 2015.

\bibitem{ding1993long}
Z.~Ding, C.~W. Granger, and R.~F. Engle.
\newblock A long memory property of stock market returns and a new model.
\newblock {\em Journal of Empirical finance}, 1(1):83--106, 1993.

\bibitem{duan1997augmented}
J.-C. Duan.
\newblock Augmented {GARCH}(p, q) process and its diffusion limit.
\newblock {\em Journal of Econometrics}, 79(1):97--127, 1997.

\bibitem{embrechts2011multivariate}
P.~Embrechts, T.~Liniger, and L.~Lin.
\newblock Multivariate {H}awkes processes: an application to financial data.
\newblock {\em Journal of Applied Probability}, 48:367--378, 2011.

\bibitem{engle1993measuring}
R.~F. Engle and V.~K. Ng.
\newblock Measuring and testing the impact of news on volatility.
\newblock {\em Journal of Finance}, 48(5):1749--1778, 1993.

\bibitem{errais2010affine}
E.~Errais, K.~Giesecke, and L.~R. Goldberg.
\newblock Affine point processes and portfolio credit risk.
\newblock {\em SIAM Journal on Financial Mathematics}, 1(1):642--665, 2010.

\bibitem{figlewski2000leverage}
S.~Figlewski and X.~Wang.
\newblock Is the leverage effect a leverage effect?
\newblock {\em Available at SSRN 256109}, 2000.

\bibitem{filimonov2015apparent}
V.~Filimonov and D.~Sornette.
\newblock Apparent criticality and calibration issues in the hawkes
  self-excited point process model: application to high-frequency financial
  data.
\newblock {\em Quantitative Finance}, (ahead-of-print):1--22, 2015.

\bibitem{french1987expected}
K.~R. French, G.~W. Schwert, and R.~F. Stambaugh.
\newblock Expected stock returns and volatility.
\newblock {\em Journal of Financial Economics}, 19(1):3--29, 1987.

\bibitem{gatheral2014volatility}
J.~Gatheral, T.~Jaisson, and M.~Rosenbaum.
\newblock Volatility is rough.
\newblock {\em arXiv preprint arXiv:1410.3394}, 2014.

\bibitem{hagan2002managing}
P.~S. Hagan, D.~Kumar, A.~S. Lesniewski, and D.~E. Woodward.
\newblock Managing smile risk.
\newblock {\em Wilmott Magazine}, pages 84--108, 2002.

\bibitem{hardiman2013critical}
S.~J. Hardiman, N.~Bercot, and J.-P. Bouchaud.
\newblock Critical reflexivity in financial markets: a hawkes process analysis.
\newblock {\em The European Physical Journal B}, 86(10):1--9, 2013.

\bibitem{haubold2011mittag}
H.~J. Haubold, A.~M. Mathai, and R.~K. Saxena.
\newblock Mittag-leffler functions and their applications.
\newblock {\em Journal of Applied Mathematics}, 2011.

\bibitem{hawkes1971point}
A.~G. Hawkes.
\newblock Point spectra of some mutually exciting point processes.
\newblock {\em Journal of the Royal Statistical Society. Series B
  (Methodological)}, pages 438--443, 1971.

\bibitem{hawkes1974cluster}
A.~G. Hawkes and D.~Oakes.
\newblock A cluster process representation of a self-exciting process.
\newblock {\em Journal of Applied Probability}, pages 493--503, 1974.

\bibitem{hendershott2006market}
T.~Hendershott and M.~S. Seasholes.
\newblock Market maker inventories and stock prices.
\newblock {\em Available at SSRN 890860}, 2006.

\bibitem{heston1993closed}
S.~L. Heston.
\newblock A closed-form solution for options with stochastic volatility with
  applications to bond and currency options.
\newblock {\em Review of financial studies}, 6(2):327--343, 1993.

\bibitem{ho1981optimal}
T.~Ho and H.~R. Stoll.
\newblock Optimal dealer pricing under transactions and return uncertainty.
\newblock {\em Journal of Financial Economics}, 9(1):47--73, 1981.

\bibitem{hull1987pricing}
J.~Hull and A.~White.
\newblock The pricing of options on assets with stochastic volatilities.
\newblock {\em Journal of Finance}, 42(2):281--300, 1987.

\bibitem{jacod2013limit}
J.~Jacod and A.~Shiryaev.
\newblock {\em Limit theorems for stochastic processes}, volume 288.
\newblock Springer Science \& Business Media, 2013.

\bibitem{jaisson2015limit}
T.~Jaisson and M.~Rosenbaum.
\newblock Limit theorems for nearly unstable hawkes processes.
\newblock {\em The Annals of Applied Probability}, 25(2):600--631, 2015.

\bibitem{jaisson2016rough}
T.~Jaisson and M.~Rosenbaum.
\newblock Rough fractional diffusions as scaling limits of nearly unstable
  heavy tailed hawkes processes.
\newblock {\em The Annals of Applied Probability}, (To appear), 2016.

\bibitem{jakubowski1989convergence}
A.~Jakubowski, J.~M{\'e}min, and G.~Pag{\`e}s.
\newblock Convergence en loi des suites d'int{\'e}grales stochastiques sur
  l'espace 1 de skorokhod.
\newblock {\em Probability Theory and Related Fields}, 81(1):111--137, 1989.

\bibitem{kurtz1991weak}
T.~G. Kurtz and P.~Protter.
\newblock Weak limit theorems for stochastic integrals and stochastic
  differential equations.
\newblock {\em The Annals of Probability}, pages 1035--1070, 1991.

\bibitem{lehalle2013market}
C.-A. Lehalle and S.~Laruelle.
\newblock {\em Market microstructure in practice}.
\newblock World Scientific, 2013.

\bibitem{lindner2009continuous}
A.~M. Lindner.
\newblock Continuous time approximations to garch and stochastic volatility
  models.
\newblock In {\em Handbook of Financial Time Series}, pages 481--496. Springer,
  2009.

\bibitem{mainardisome}
F.~Mainardi.
\newblock On some properties of the {M}ittag-{L}effler function.
\newblock {\em arXiv preprint arXiv:1305.0161}.

\bibitem{mathai2008special}
A.~M. Mathai and H.~J. Haubold.
\newblock {\em Special functions for applied scientists}.
\newblock Springer, 2008.

\bibitem{mytnik2015uniqueness}
L.~Mytnik and T.~S. Salisbury.
\newblock Uniqueness for {V}olterra-type stochastic integral equations.
\newblock {\em arXiv preprint arXiv:1502.05513}, 2015.

\bibitem{nelson1990arch}
D.~B. Nelson.
\newblock {ARCH} models as diffusion approximations.
\newblock {\em Journal of Econometrics}, 45(1):7--38, 1990.

\bibitem{nelson1991conditional}
D.~B. Nelson.
\newblock Conditional heteroskedasticity in asset returns: A new approach.
\newblock {\em Econometrica}, pages 347--370, 1991.

\bibitem{revuz1999continuous}
D.~Revuz and M.~Yor.
\newblock {\em Continuous martingales and Brownian motion}, volume 293.
\newblock Springer Science \& Business Media, 1999.

\bibitem{rodriguez2012revisiting}
M.~J. Rodr{\'\i}guez and E.~Ruiz.
\newblock Revisiting several popular garch models with leverage effect:
  Differences and similarities.
\newblock {\em Journal of Financial Econometrics}, 10(4):637--668, 2012.

\bibitem{stein1991stock}
E.~M. Stein and J.~C. Stein.
\newblock Stock price distributions with stochastic volatility: {A}n analytic
  approach.
\newblock {\em Review of Financial Studies}, 4(4):727--752, 1991.

\bibitem{tayal2012measuring}
R.~Tayal and S.~Thomas.
\newblock Measuring and explaining the asymmetry of liquidity.
\newblock {\em Available at SSRN 2239492}, 2012.

\bibitem{veraar2012stochastic}
M.~Veraar.
\newblock The stochastic fubini theorem revisited.
\newblock {\em Stochastics An International Journal of Probability and
  Stochastic Processes}, 84(4):543--551, 2012.

\bibitem{wang2014estimation}
C.~D. Wang and P.~A. Mykland.
\newblock The estimation of leverage effect with high-frequency data.
\newblock {\em Journal of the American Statistical Association},
  109(505):197--215, 2014.

\bibitem{wu2002generalized}
G.~Wu and Z.~Xiao.
\newblock A generalized partially linear model of asymmetric volatility.
\newblock {\em Journal of Empirical Finance}, 9(3):287--319, 2002.

\bibitem{zakoian1994threshold}
J.-M. Zakoian.
\newblock Threshold heteroskedastic models.
\newblock {\em Journal of Economic Dynamics and control}, 18(5):931--955, 1994.

\end{thebibliography}

\end{document}